\documentclass[floatfix,twocolumn,prb,amsmath,amssymb]{revtex4}

\usepackage{graphicx}
\usepackage{dcolumn}
\usepackage{bm}
\usepackage[section]{placeins}

\def\R{{\mathbb R}}

\def\d{{\rm d}}
\def\eps{{\varepsilon}}

\begin{document}

\title{Landau--Zener type surface hopping algorithms}

\author{Andrey K. Belyaev}
\email{belyaev@herzen.spb.ru}
\affiliation{Department of Theoretical Physics, Herzen University, St. Petersburg 191186, Russia}

\author{Caroline Lasser}
\email{classer@ma.tum.de}
\affiliation{Zentrum Mathematik, Technische Universit\"at M\"unchen, Germany}

\author{Giulio Trigila}
\email{trigila@ma.tum.de}
\affiliation{Zentrum Mathematik, Technische Universit\"at M\"unchen, Germany}

\date{\today}

\begin{abstract}
A class of surface hopping algorithms is studied comparing two recent Landau-Zener (LZ) formulas for the probability of nonadiabatic transitions. One of the formulas requires a diabatic representation of the potential matrix while the other one depends only on the adiabatic potential energy surfaces. For each classical trajectory, the nonadiabatic transitions take place only when the surface gap attains a local minimum. Numerical experiments are performed with deterministically branching trajectories and with probabilistic surface hopping. The deterministic and the probabilistic approach confirm the affinity of both the LZ probabilities, as well as the good approximation of the reference solution computed by solving the Schr\"{o}dinger equation via a grid based pseudo-spectral method. Visualizations of position expectations and superimposed surface hopping trajectories with reference position densities illustrate the effective dynamics of the investigated algorithms.
\end{abstract}

\maketitle

\section{Introduction}

A great variety of 
physical processes and chemical reactions occurs due to 
nonadiabatic transitions between adiabatic electronic states, often mediated by conical intersections \cite{Nakamura:2002, DYK:2004, Bersuker:2006, DYK:2011, DomckeYakony:2012}. 
Nonadiabatic transitions are of quantum nature and, in general, should be described with quantum mechanical theory.  


While for small molecular systems the nonadiabatic effects can indeed be investigated in detail with quantum mechanical methods, for larger systems these methods are computationally too expensive. Example of such systems are  biomolecules, atomic and molecular clusters, molecular complexes and condensed matter.

For this reason, more approximate classical or semiclassical computational methods are becoming an important alternative because their lower computational cost scales more favorably with the system size.  
Moreover, these methods can provide intuitive insight into the dynamics of a chemical reaction \cite{Tully:1998}. 
Particularly interesting for practical purposes are mixed quantum--classical approaches which treat the electronic motion quantum-mechanically and the nuclear motion classically.

Several quasi-classical methods exist for the treatment of the nonadiabatic nuclear dynamics. 
Well-known examples are the semiclassical initial-value representation (IVR) \cite{Miller:1970, Marcus:1972, KreekMarcus:1972, Miller:2001}, 
the Ehrenfest dynamics method \cite{McLachlan:1964, MeyerMiller:1979, Micha:1983, Kirson-etc:1984, Sawada-etc:1985}, 
the frozen Gaussian wave-packet method \cite{Heller:1991}, 
the propagation of classical trajectories with surface hopping \cite{BjerreNikitin:1967, TullyPreston:1971, StineMuckerman:1976, Kuntz-etc:1979, BlaisTruhlar:1983, Tully:1990, HammesSchiffer-Tully:1994, Voronin-etc:1998, Thiel-etc:2008, FermanianLasser:2008,LasserSwart:2008,MuellerStock:1997}, 
as well as the multiple-spawning wave-packet method \cite{Ben-NunMartinez:1998, Martinez-etc:2000, Ben-NunMartinez:2002, Martinez-etc:2012}; 
see the leading Perspective\cite{Tully:2012} of the special issue dedicated to nonadiabatic nuclear dynamics. Also the mathematical literature 
provides rigorous analytical results on nonadiabatic nuclear dynamics\cite{Hagedorn:1994,HagedornJoye:1998,FermanianGerard:2002,LasserTeufel:2005,FermanianLasser:2008a}. 

One of the most widely used mixed quantum-classical approaches for simulating nonadiabatic dynamics is the classical trajectory surface-hopping method with its many variants.  
To the best of our knowledge, the combination of classical trajectories and surface hopping was first introduced by \citet{BjerreNikitin:1967}, though with reduced dimensionality. 
They proposed to branch a classical trajectory into two trajectories after traversing a nonadiabatic region which had to be specified beforehand. 
The more systematic classical trajectory surface-hopping approach was proposed by \citet{TullyPreston:1971} based on the deterministic (``ants'') procedure or/and on the probabilistic (``anteater'') method. 
In the latter, classical trajectories remain unbranched and a random decision is made whether to hop or not, depending on a hopping probability. 
In both papers, nonadiabatic transition probabilities were estimated in an approximate way within the Landau-Zener (LZ) model \cite{Landau:1932a, Landau:1932b, Zener:1932}, with parameters calculated beforehand. 
\citet{TullyPreston:1971} have also used semiclassical methods to investigate the hopping probability and to prove LZ model usage.  
Later, \citet{Kuntz-etc:1979} proposed a probabilistic approach in which hopping points were not specified beforehand but determined during the trajectory propagation, based on a maximum of the nonadiabatic time-derivative coupling matrix element. 
Approaches to localize nonadiabatic regions by a local minimum for an adiabatic splitting have been proposed by 
 \citet{MillerGeorge:1972} as well as \citet{StineMuckerman:1976}. 
In contrast to these approaches, where nonadiabatic transitions are localized, \citet{Tully:1990} proposed the fewest-switches approach, which extended the classical trajectory surface-hopping method to an arbitrary number of states and to situations in which transitions can occur anywhere, not just at localized regions. 
This is achieved by a solution of the time-dependent Schr\"odinger equation along classical trajectories in combination with the probabilistic fewest-switches algorithm that decides at each integration time step whether to switch the electronic state. 
Since then, many variants of the classical trajectory surface-hopping approaches have been proposed and applied to different physical phenomena and processes. 
The main differences between different classical trajectory surface-hopping versions are in two features: 
(i) how a nonadiabatic region (a seam) is defined, 
and (ii) when and how a hopping probability is determined. 
The present paper is addressed to these questions in connection with a conical intersection case. 

The simplicity of the classical trajectory surface-hopping technique renders it attractive for the study of high-dimensional quantum systems which are difficult or unreachable for quantum treatments. 
Today, trajectory surface-hopping calculations are widely employed in the context of so-called {\em ab initio} molecular dynamics, that is, the forces for the trajectory calculation and the nonadiabatic couplings are computed "on-the-fly" with {\em ab initio} or semiempirical electronic-structure methods, see, e.g., Ref.~\cite{NewtonX}. 
Nevertheless, many surface-hopping methods have been derived and tested for one- or two-dimensional cases. 
In the present paper, we treat a two-dimensional two-state model for studying nonadiabatic transitions in the vicinity of a conical intersection by different classical trajectory approaches.  

A classical trajectory surface-hopping simulation of nonadiabatic dynamics involves the following steps: 
(i) sampling of the initial condition,  
(ii) performing classical trajectory calculations on multi-dimensional adiabatic potential energy surfaces (PES), 
(iii) accounting for nonadiabatic effects through surface hopping according to specified criteria, 
and (iv) evaluation of the observables of interest from the ensemble of trajectories.

The important feature distinguishing different surface-hopping approaches is the way of calculating nonadiabatic transition probabilities. 
There are several solutions to this problem, many of them based on the LZ model, see, e.g., \cite{BjerreNikitin:1967, TullyPreston:1971, Kuntz-etc:1979, Voronin-etc:1998, Thiel-etc:2008, FermanianLasser:2008, BelyaevLebedev:2011}.
Although the LZ model provides the simple formula for a nonadiabatic transition probability (see below), it is formulated as a one-dimensional problem in a two-state diabatic representation. 
In practical applications to polyatomic systems however, nonadiabatic transitions occur in a multi-dimensional space and quantum-chemical data are usually provided in an adiabatic representation, for example, for an on-the-fly study. 
Moreover, often only adiabatic PESs are available, not nonadiabatic couplings. 
As is well known, in contrast to adiabatic states, diabatic states are not uniquely defined, and diabatic representations obtained by the same procedure in two-state and in multiple-state cases may deviate substantially \cite{Belyaev-etc:2010}. 
Lastly, a determination of LZ parameters is often troublesome in practical applications of the conventional LZ formula.

Two novel formulas have recently been proposed for nonadiabatic transition probabilities within the LZ model: the diabatic multi-dimensional formula \cite{FermanianLasser:2008} and the adiabatic-potential-based transition probability formula \cite{BelyaevLebedev:2011} adapted for classical trajectory surface-hopping studies. 
The former was derived when mathematically analysing effective dynamics through conical intersections\cite{FermanianLasser:2008a} and tested on the two-state three-mode model of pyrazine \cite{LasserSwart:2008} by means of the single switch classical trajectory surface-hopping algorithm, while the latter was applied to inelastic multi-channel atomic collisions by means of the branching classical trajectory \cite{BelyaevLebedev:2011} and the branching probability current algorithms \cite{Belyaev:2013}. 
The formula derived in Ref.~\cite{BelyaevLebedev:2011} is easy implemented in practice as it only requires the information about adiabatic potentials (see below). 
It should be mentioned that \citet{ZhuNakamura:2001} have derived the formula 
for the LZ transition probability written in terms of several parameters that are expressed via adiabatic potentials, but the Zhu-Nakamura formula is different from the adiabatic-potential-based formula \cite{BelyaevLebedev:2011}. 
\citet{TullyPreston:1971, StineMuckerman:1976, Voronin-etc:1998}  have calculated transition probabilities by means of the conventional LZ formula with diabatic LZ parameters determined from adiabatic potentials along a trajectory. 
Their approaches are also different from the one of Ref.~\cite{BelyaevLebedev:2011}. 
The adiabatic-potential-based formula has been applied so far to nonadiabatic transitions in atomic collisions\cite{BelyaevLebedev:2011, Belyaev:2013}. 

Thus, the main goal of the present work is to study different versions of classical trajectory surface-hopping algorithms based on the novel formulas for nonadiabatic transition probabilities within the Landau-Zener model in their applications to a two-state two-dimensional model for a conical intersection. 
In addition, we test probabilistic versus deterministic versions of the algorithm and study simulation performance for several consecutive nonadiabatic transition phases. We also explore the possibilities of visualizing nonadiabatic dynamics by surface-hopping simulations.

\section{Surface hopping with Wigner functions}

Molecular quantum motion is governed by the Schr\"odinger operator
$$
H_{\rm mol} = T + T_{\rm el} + V_{\rm el} + V_{\rm nuc} + V_{\rm attr}, 
$$
where $V_{\rm el}$ and $V_{\rm nuc}$ denote electronic and nuclear repulsion, respectively, and $V_{\rm attr}$ the attraction between electrons and nuclei. By a rescaling of the nuclear coordinates, we can assume that all nuclei have identical mass~$m$. Then, the kinetic energy operators are
$$
T = -\frac{\hbar^2}{2m} \Delta_{\rm nuc},\quad
T_{\rm el} = -\frac{\hbar^2}{2m_{\rm el}}\Delta_{\rm el},\quad
$$
where $\Delta_{\rm nuc}$ and $\Delta_{\rm el}$ denote Laplacians with respect to the nuclear and electronic coordinates. Moving to atomic units ($\hbar=m_{\rm el}=e=1$) and defining 
$$
\eps=1/\sqrt{m}, 
$$  
the molecular Hamiltonian rewrites as 
$$
H_{\rm mol} = -\frac{\eps^2}{2}\Delta_{\rm nuc} 
-\frac{1}{2}\Delta_{\rm el} + V_{\rm el} + V_{\rm nuc} + V_{\rm attr}
$$

Let $H_{\rm el}(q) = -\frac{1}{2}\Delta_{\rm el} + V_{\rm el} + V_{\rm nuc}(q) + V_{\rm attr}(\cdot,q)$ be the electronic Hamiltonian for a given position $q\in\R^d$ of the nuclei. Let 
$$
U^\pm(q)\in\sigma(H_{\rm el}(q))
$$ 
be two adiabatic potential energy surfaces (PES), that is, two eigenvalues of the electronic Hamiltonian. We assume that each eigenvalue is of  multiplicity one and that both are well separated from the rest of the electronic spectrum. Then, by time-dependent Born--Oppenheimer theory \cite{Hagedorn:1980, SpohnTeufel:2001}, the effective nuclear quantum motion is governed by the time-dependent nuclear Schr\"odinger equation 
\begin{equation}\label{eq:schro}
i\eps\partial_t \psi_t= (T + V)\psi_t,
\end{equation} 
where the potential $V$ takes values in the  real symmetric $2\times 2$ matrices and is given in a global diabatic representation as
$$
V(q) = \begin{pmatrix}v_{11}(q) & v_{12}(q)\\ v_{12}(q) & v_{22}(q)\end{pmatrix}.
$$ 
We note that by the definition of a diabatic matrix, the potential energy surfaces are its eigenvalues. In the present work, we assume the existence of a conical intersection, that is, 
$$
\{q\mid U^+(q)=U^-(q)\}
$$ 
is a manifold of codimension two of the nuclear configuration space. Then, one has to account for nonadiabatic transitions between the eigenspaces associated with the potential energy surfaces. 

For notational convenience, we write the diabatic matrix as the sum of a centric dilation and its trace-free part,  
$$
V(q) = v_0(q) + \begin{pmatrix} v_1(q) & v_2(q)\\ v_2(q) & -v_1(q)\end{pmatrix},
$$
and express the two potential energy surfaces as 
$$
U^\pm(q) = v_0(q)\pm \sqrt{v_1(q)^2 + v_2(q)^2}.
$$
Their gap is denoted by 
$$
Z(q) = U^+(q)-U^-(q).
$$  
The corresponding eigenvectors of $V(q)$ satisfy $V(q)\chi^\pm(q) = U^\pm(q)\chi^\pm(q)$. 
They are uniquely determined up to a phase and are singular at conical intersection points. Our choice for the phase is
$$
\chi^+(q) = \begin{pmatrix}\cos(\alpha(q))\\ \sin(\alpha(q))\end{pmatrix},\quad
\chi^-(q) = \begin{pmatrix}-\sin(\alpha(q))\\ \cos(\alpha(q))\end{pmatrix},
$$
with mixing angle $\alpha(q) = \tfrac12 \arctan(v_2(q)/v_1(q))$.

\subsection{The observables}
We write the wave function at time $t$ as a linear combination of the eigenvectors  
$\psi_t = \psi_{t}^+ \chi^+ + \psi_{t}^- \chi^-$ 
with scalar-valued functions $\psi_t^+$ and $\psi_t^-$. We use the Wigner functions of 
the scalar components 
\begin{eqnarray*}
\lefteqn{W(\psi_{t}^\pm)(q,p) =}\\
&&(2\pi\eps)^{-d} \int e^{iy\cdot p/\eps} \psi_t^\pm(q-\tfrac12 y) \psi_t^\pm(q+\tfrac12 y)^* \d y,
\end{eqnarray*}
which map phase space points $(q,p)\in\R^{2d}$ to the real numbers. The $\eps$-scaling of the Wigner function allows the direct relation to the position, momentum, and kinetic energy operators,
$$
\hat q = q,\qquad
\hat p = -i\eps\nabla,\qquad
T = -\frac{\eps^2}{2}\Delta = \frac12 \sum_{j=1}^d \hat p_j^2
$$ 
for the nuclear degrees of freedom. Indeed, up to normalizing factors, we obtain the corresponding expectation values 
of the upper and the lower adiabatic surface as
\begin{eqnarray*}
\langle \psi_t^\pm\mid \hat q \mid\psi_t^\pm\rangle  &=& 
\int q\, W(\psi_t^\pm)(q,p) \d(q,p),\\
\langle \psi_t^\pm\mid \hat p \mid\psi_t^\pm\rangle  &=& 
\int p\, W(\psi_t^\pm)(q,p) \d(q,p),\\
\langle \psi_t^\pm\mid T \mid\psi_t^\pm\rangle  &=& 
\int \tfrac12|p|^2\, W(\psi_t^\pm)(q,p) \d(q,p).
\end{eqnarray*}
More general expectation values for the Weyl quantization $\hat A$ of a phase space function $A$ are accordingly written as
$$
\langle \psi_t^\pm\mid \hat A \mid\psi_t^\pm\rangle  = 
\int A(q,p) W(\psi_t^\pm)(q,p) \d(q,p).
$$
These are the observables whose dynamics can be approximated by surface hopping algorithms. 

Cross-term quantities like 
$\langle \psi^\pm_t\mid \hat A\mid \psi^\mp_t\rangle$, 
which require relative phase information of the upper and the lower surface components cannot be obtained, 
since the  Wigner functions $W(\psi^+_t)$ and $W(\psi^-_t)$ determine the functions $\psi_t^+$ and $\psi_t^-$  only up to a global phase factor.

\subsection{The general algorithmic scheme}
The class of surface hopping algorithms to be investigated is determined by the following steps: 

(i) Sampling of the initial condition: We choose phase space points $(q_1^\pm,p_1^\pm),\ldots,(q_{N_0}^\pm,p_{N_0}^\pm)$ so that
$$
\langle \psi_0^\pm\mid \hat A \mid\psi_0^\pm\rangle \approx \frac{1}{N_0} \sum_{j=1}^{N_0} A(q_j^\pm,p_j^\pm) 
$$
for the observables $A$ of interest. This is achieved by Monte Carlo or Quasi-Monte Carlo sampling of the initial Wigner functions $W(\psi^\pm_0)$. We note that an unrefined sampling of the initial Husimi functions deteriorates the approximation\cite{KubeLasserWeber:2009,KellerLasser:2013}.

(ii) Classical trajectory calculations: The chosen phase space points are evolved along the trajectories of the corresponding classical Hamiltonian system
$$
\dot q = p,\qquad \dot p = -\nabla U^\pm(q).
$$
Since the observables of interest are computed by phase space averaging, these classical equations of motion should be discretized symplectically as e.g. by the St\"ormer--Verlet method or by higher order symplectic 
Runge--Kutta schemes\cite{LasserRoeblitz:2010}.

(iii) Surface hopping: Whenever the eigenvalue gap becomes minimal along an individual classical trajectory a nonadiabatic transition occurs. Let $t\mapsto(q(t),p(t))$ be a classical trajectory associated with the upper or the lower surface. Whenever the function $t\mapsto Z(q(t))$ attains a local minimum, a transition to the other eigenspace is performed according to a Landau--Zener transition probability. In the following sections \S\ref{sec:DifferentLZ} and \S\ref{sec:TransitionSchemes}, different Landau-Zener formulas and transition schemes will be discussed in more detail.

(iv) Evaluation of the observables: At some time~$t$, the surface hopping algorithm has resulted in phase space points $(q_1^\pm(t),p_1^\pm(t)),\ldots,(q_{N_t}^\pm(t),p_{N_t}^\pm(t))$. Then, the expectation values of interest are approximated according to
\begin{equation}\label{eq:GridEx}
\langle \psi_t^\pm\mid \hat A \mid\psi_t^\pm\rangle \approx \sum_{j=1}^{N_t} A(q_j^\pm(t),p_j^\pm(t)) w_j^\pm(t), 
\end{equation}
where the individual weight $w_j^\pm(t)$ depends on the initial sampling and the employed transition scheme, see \S\ref{sec:TransitionSchemes}.

\section{Two Landau--Zener probabilities}\label{sec:DifferentLZ}

We compare two recent formulas for nonadiabatic transition probabilities. Both of them are applied whenever the eigenvalue gap becomes minimal along an individual classical trajectory $t\mapsto(q(t),p(t))$, that is, when the function $t\mapsto Z(q(t))$ attains a local minimum. We denote corresponding critical times and phase space points by $t_c$ and $(q_c,p_c)$ respectively. 

The first formula is a multi-dimensional Landau--Zener formula derived from a global diabatic representation of the potential matrix\cite{FermanianLasser:2008a,FermanianLasser:2008}
\begin{equation}\label{eq:dia}
P_d^{\rm LZ} = \exp\!\left(-\frac{\pi}{\eps} \;\frac{Z(q_c)^2}{4|\d v(q_c)p_c|}\right),
\end{equation}
where $\d v(q)$ denotes the $2\times d$ gradient matrix of the vector $v(q) = (v_1(q),v_2(q))$ defining the trace-free part of the diabatic potential matrix $V(q)$. 
The second formula is the purely gap and trajectory based, adiabatic formula\cite{BelyaevLebedev:2011}
\begin{equation}\label{eq:ad}
P_a^{\rm LZ} = \exp\!\left( -\frac{\pi}{2\eps} \;\sqrt{\frac{Z(q_c)^3}{\frac{d^2}{dt^2} Z(q(t))\mid_{t=t_c}}}\right)
\end{equation}
Contrary to the diabatic formula, the building blocks of the adiabatic one are accessible also in cases when a diabatic potential matrix is missing, which is often the case for the simulation of polyatomic systems. A simple calculation reveals the connection between the two LZ formulas: 
Depending on the potential energy surface $U^\pm$ guiding the classical motion, we have
$$
\frac12\sqrt{\frac{Z(q_c)^3}{\frac{d^2}{dt^2} Z(q(t))\mid_{t=t_c}}} = \frac{Z(q_c)^2}{4\sqrt{|\d v(q_c)p_c|^2 + v(q_c)\cdot w^\pm(q_c,p_c)}}
$$
with
$$
w^\pm(q,p) = \begin{pmatrix}D^2v_1(q)p\cdot p\\ D^2v_2(q)p\cdot p\end{pmatrix}-\d v(q)\nabla U^\pm(q)
$$ 
where $D^{2}v(q)$ denotes the Hessian matrix of $v(q)$.
Since $|v(q)\cdot w^\pm(q,p)| \le \tfrac12 |w^\pm(q,p)|\, Z(q)$, the difference between the two formulas is dominated by the gap size and negligible for trajectories with small minimal gap. Nevertheless, we observe a notable difference. 
The diabatic formula has the same functional form for transitions originating from the upper or the lower surface, while the adiabatic formula implicitly depends on the surface with which the hopping trajectory is associated.

\section{Transition schemes}\label{sec:TransitionSchemes}
One can algorithmically interprete nonadiabatic transitions with LZ probabilities either in a deterministic way with branching trajectories or probabilistically with surface hopping trajectories. The probabilistic method is appealing, since it is less costly from the computational point of view.
The deterministic branching process has been mathematically analysed\cite{FermanianLasser:2008a} and has been proven to be asymptotically correct in the semiclassical limit $\eps\to0$, provided that at the same time $t_c$ and in the same phase point $(q_c,p_c)$ only upper or lower surface trajectories initiate nonadiabatic transitions. This restriction is due to the neglect of relative phase information for the upper and the lower wavefunctions $\psi^+_t$ and $\psi^-_t$. The numerical experiments presented later confirm the mathematical assessment of the algorithm's properties.

\subsection{Deterministic transitions}
For the deterministic branching process\cite{FermanianLasser:2008,BelyaevLebedev:2011}, at a critical point $(q_{c},p_{c})$ of minimal gap a trajectory splits into two, and a new branch is created on the other surface. The weight of the new trajectory is equal to the old weight multiplied by the Landau--Zener probability $P^{\rm LZ}$, while the weight of the trajectory remaining on the same surface is multiplied by $1-P^{\rm LZ}$. At time $t$, we are left with a certain number of classical trajectories distributed along the upper and the lower surface. Let us indicate with $N_t$ the number of trajectories on the upper surface and with $w_1^+(t),\ldots,w_{N_t}^+(t)$ the corresponding weights. Each weight $w_j^+(t)$ is the product of the initial sampling weight $1/N_0$ and a certain number of Landau--Zener probabilities. Corresponding expectation values are computed according to (\ref{eq:GridEx}). We note that the number of trajectories may rapidly increase as time evolves, demanding  more memory storage and 
increasing computational costs.  

\subsection{Probabilistic transitions}
In constrast to the deterministic method, the probabilistic version of the surface hopping algorithm keeps the number of trajectories constant during all the simulation time. In this case, once a classical trajectory attains a local gap minimum, we compute the LZ probability $P^{\rm LZ}$ and compare it with a pseudo random number $\xi$ generated from a uniform distribution on $[0,1]$. If $\xi \leq P^{\rm LZ}$ the trajectory hops on the other surface, otherwise it continues along the same surface. The weights in the expectation summation (\ref{eq:GridEx}) are all equal to $1/N_0$. 

\subsection{Momentum adjustment}
In both methods described above, a choice has to be made regarding the point in phase space at which a trajectory appears on the other energy surface at the moment of a nonadiabatic transition. 
For clarity, let us consider a transition from the upper to the lower surface. We denote with $(q^{+},p^{+})$ the point on the upper surface, in which the trajectory attains a local gap minimum, and with $(q^{-},p^{-})$ the point on the lower surface, in which a new trajectory is initiated. Our choice is $q^{-}=q^{+}=q_c$ for the position, and we rescale the momentum according to $p^{-}=kp^{+}$ with $k>0$ to ensure conservation of energy. The value of $k$ is computed by simply imposing
$$
\tfrac12|p^{+}|^2 + U^+(q_c)= \tfrac12|kp^{+}|^2 + U^-(q_c),
$$
leading to $k=\sqrt{1+2Z(q_{c})/|p^{+}|^{2}}$.
Analogously for transitions from the lower to the upper surface, we have 
$k=\sqrt{1-2Z(q_{c})/|p^{-}|^{2}}$, where we neglect the transition if the trajectory does not have enough kinetic energy to compensate the difference in the potential energy. 

This particularly simple momentum adjustment seems natural from the classical trajectory point of view, since it treats each newly generated trajectory as a continuation of its generator, while ensuring that both trajectories have the same classical energy.  Moreover, it only depends on the adiabatic surfaces and their gap. With respect to the direction of the momentum adjustment, the literature contains various other, more complicated choices, such as the normal direction to a predefined surface of (avoided) intersection\cite{MillerGeorge:1972,StineMuckerman:1976} or the direction of the nonadiabatic coupling vector\cite{Tully:1990} defined as $(\chi^+(q)\cdot \partial_1 \chi^-(q),\ldots,\chi^+(q)\cdot\partial_d\chi^-(q))$.

\section{Numerics}
We now compare the two Landau--Zener transition probabilities for the Schr\"{o}dinger equation~(\ref{eq:schro}) associated with a linear Jahn--Teller matrix 
$$
V(q) = \gamma |q|^2 + \begin{pmatrix}q_1&q_2\\q_2&-q_1\end{pmatrix},
$$
where the quadratic term confines the motion around the conical intersection located at the origin. 
The strength of the confinement is chosen to be $\gamma=3$, while for the semiclassical parameter we set $\eps=0.01$. These values are comparable with those obtained by ab initio electronic structure calculations for the triangular silver molecule\cite{Garcia:2005}. 

The initial wave packet is localized entirely on the upper surface and is given by 
$\psi_{0}= \psi_{0}^+ \chi^+$ where
$$
\psi_{0}^+(q)=(\pi \eps)^{-1/2}\exp\left(-\tfrac{1}{2\eps}|q-q_{0}|^{2} 
\right).
$$
The initial position center is $q_{0}=(5\sqrt{\eps},0.5\sqrt{\eps})$, so that the wave packet is localized close to the conical intersection. The final simulation time $t_f=5.34$ roughly corresponds to $129.3$ fs and allows the wavefunction to pass the conical intersection four times.  


As previously developed for the computation of expectation values via the Wigner function \cite{LasserRoeblitz:2010}, the initial Wigner function 
\begin{equation}\label{eq:Wo}
W(\psi_{0}^+)(q,p) = (\pi\eps)^{-2} \exp\!\left(-\tfrac1\eps(|q-q_0|^2 + |p|^2)\right)
\end{equation}
is sampled by a quasi-Monte Carlo technique so that
\begin{eqnarray*}
\langle \psi_0^+\mid \hat A \mid \psi_0^+\rangle &=& \int A(q,p) W(\psi_0^+)(q,p) d(q,p)\\
  &\approx & \frac{1}{N_0} \sum_{j=1}^{N_0} A(q_j^+,p_j^+)
\end{eqnarray*}
for the observables $\hat A$ of interest. We have used $N_0=1296$ Halton points 
$(q_1^+,p_1^+),\ldots,(q_{N_0}^+,p_{N_0}^+)$, 
 which deterministically approximate the uniform distribution on the unit cube $[0,1)^{2d}$,  and have mapped them by the inverse of the cumulative distribution function of $2d$ one-dimensional Gaussian distributions to approximate the $2d$-dimensional Gaussian 
distribution given in equation~(\ref{eq:Wo}). The corresponding convergence rate for the approximation of expectation values scales as $(\log N_{0})^{2d}/N_{0}$ compared to the $1/\sqrt{N_{0}}$ scaling characterizing plain Monte Carlo.

The numerical integration of the classical trajectories is implemented using a 4th order symplectic Runge-Kutta time-stepping method, while, in order to estimate the second derivative $d^{2}Z(q(t))/dt^2 $ for the evaluation of the adiabatic LZ probability, a 4th order accurate central finite difference scheme is used. 

The simulations presented below are performed both in the deterministic and the probabilistic setting. For each numerical experiment, we compute the values of the surface populations and the expected values of the momentum and position, that is,
$\langle\psi^\pm_t\mid \hat A\mid \psi^\pm_t\rangle$
for $A=1$, $A=p$ and $A=q$, respectively. These expected values are compared with reference solutions computed by solving the Schr\"{o}dinger equation (\ref{eq:schro}) via a numerically converged Strang splitting scheme using the fast Fourier transform for the computation of the Laplacian. 
This grid-based reference solution $\psi_t^{\rm ref}$ approximates\cite{KubeLasserWeber:2009}  the solution~$\psi_t$ of the Schr\"odinger equation (\ref{eq:schro}) with an  accuracy of $\langle \psi_t-\psi_t^{\rm ref}\mid \psi_t-\psi_t^{\rm ref}\rangle\approx 10^{-12}$.

\subsection{Time evolution}

\begin{figure}[h]
    \resizebox{85mm}{!}{\includegraphics{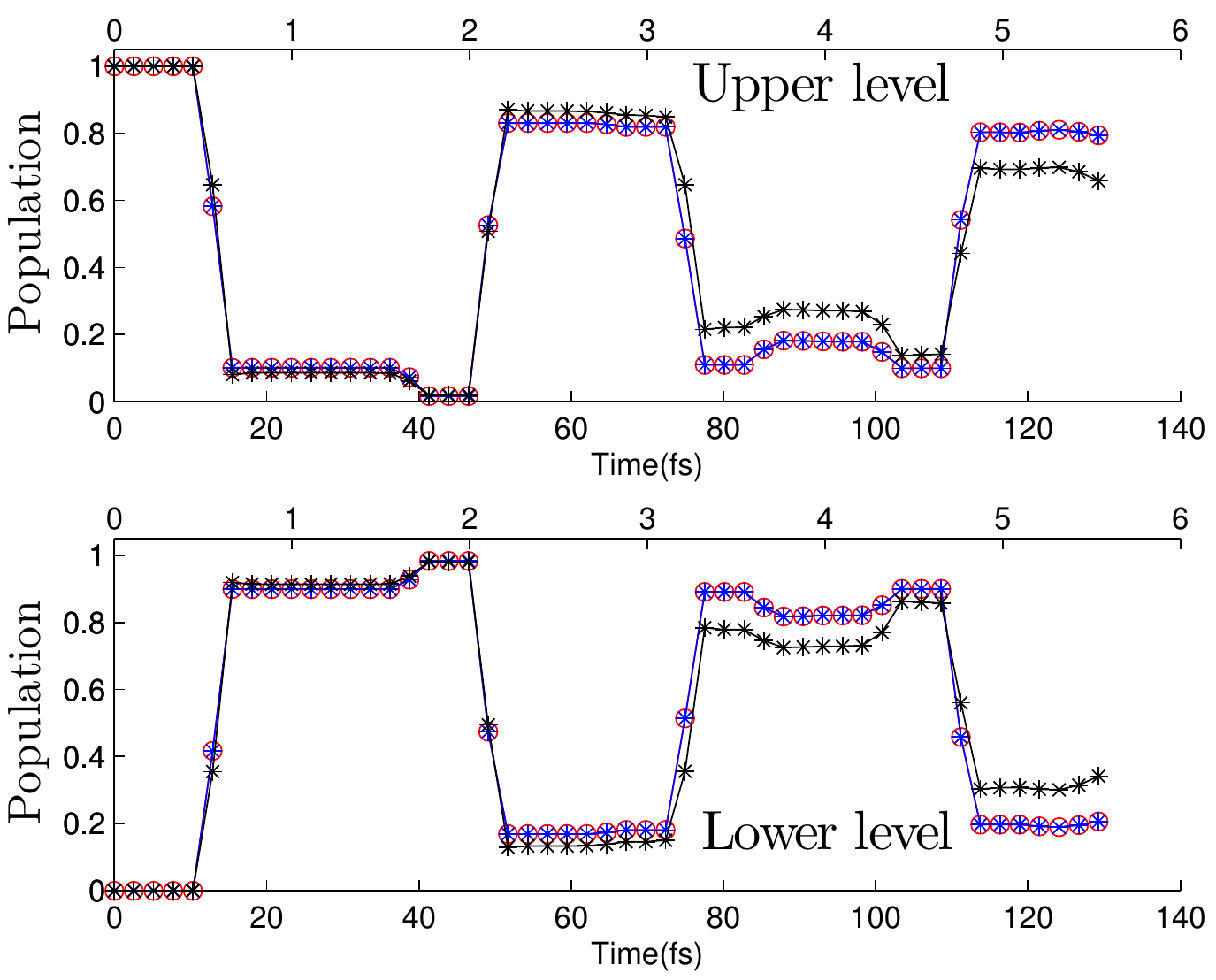}}     
    \caption{Population of the upper and lower surface, respectively. The blue markers refer to the simulation obtained using $P^{\rm LZ}_{d}$ while the red markers    
    refer to $P^{\rm LZ}_{a}$; the black curve represents the reference solution. The label on the $x$ axis at the bottom of each panel indicates the time in femtoseconds, while the axis located on top of the panel indicates the time in units that are consistent with the equation (\ref{eq:schro}).}
    \label{fig:Pop}
\end{figure}

We compare the time evolution of the above mentioned observables when using the two different LZ formulas for nonadiabatic transition  probabilities. 
In Fig.~\ref{fig:Pop} we show the population of the upper and lower surfaces calculated in the deterministic setting. As suggested by our previous analysis, the curves associated with probabilities computed by the two different LZ formulas are almost indistinguishable and in good agreement with the reference. The slight deterioration of both surface hopping approximations after the third and fourth nonadiabatic passage (around time $=80$fs and $=100$fs, respectively) are due to unresolved interference effects between the upper and the lower wave packet components. This effect is also visible 
in Fig.~\ref{fig:Error} later on.

\begin{figure*}[h]
    \begin{tabular}{llll}                                                                                      
      \resizebox{65mm}{!}{\includegraphics{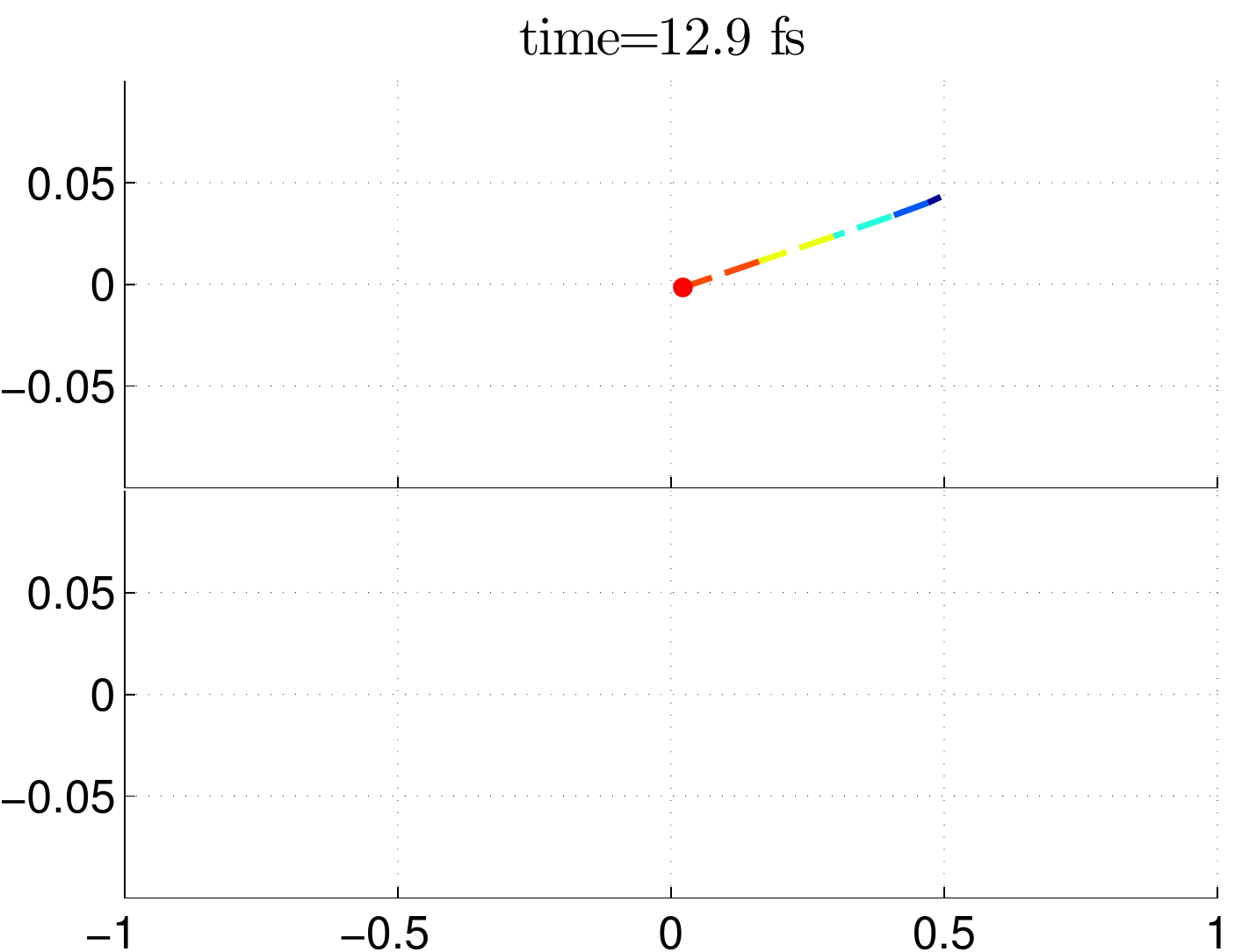}} &
      \resizebox{65mm}{!}{\includegraphics{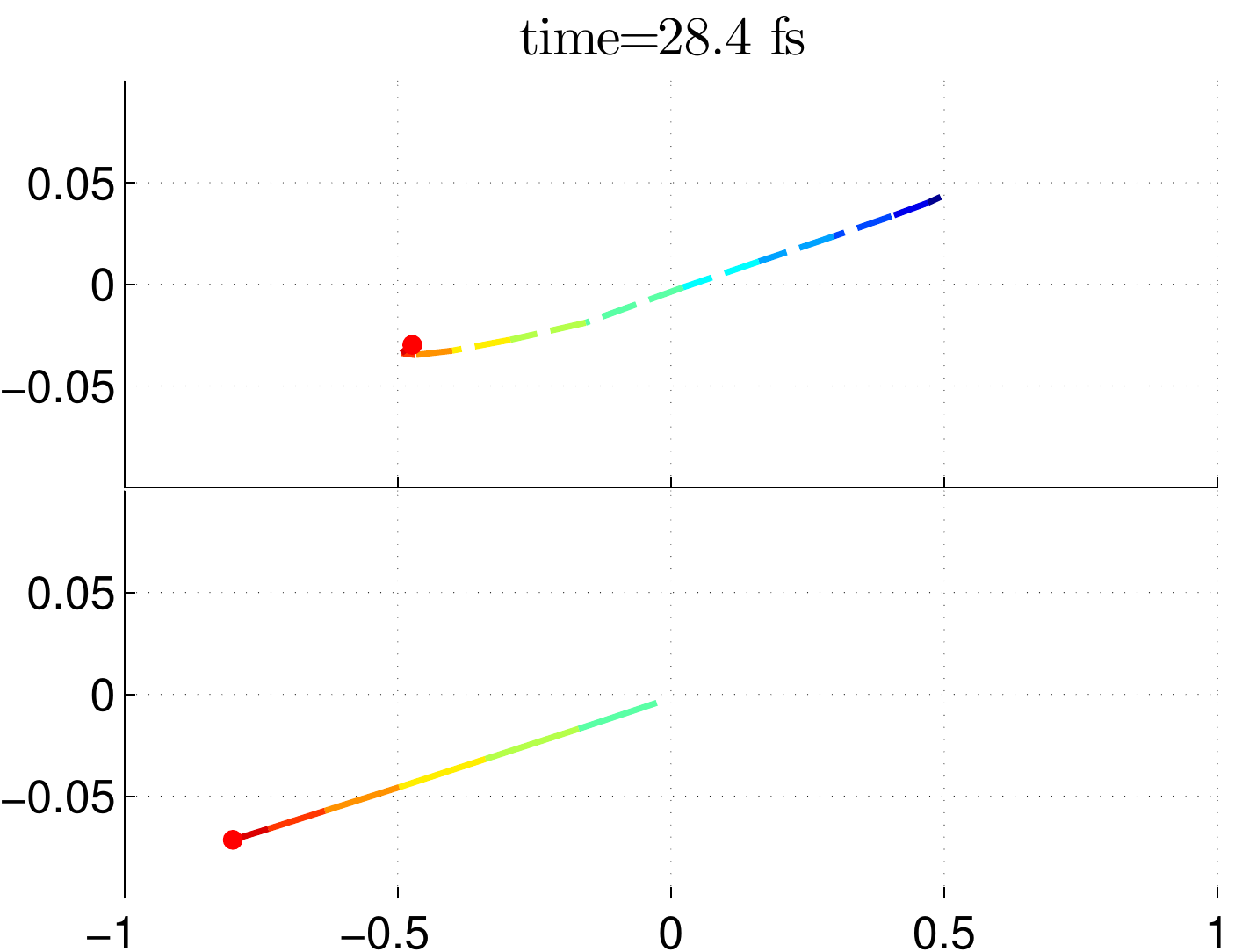}} \\
       \resizebox{65mm}{!}{\includegraphics{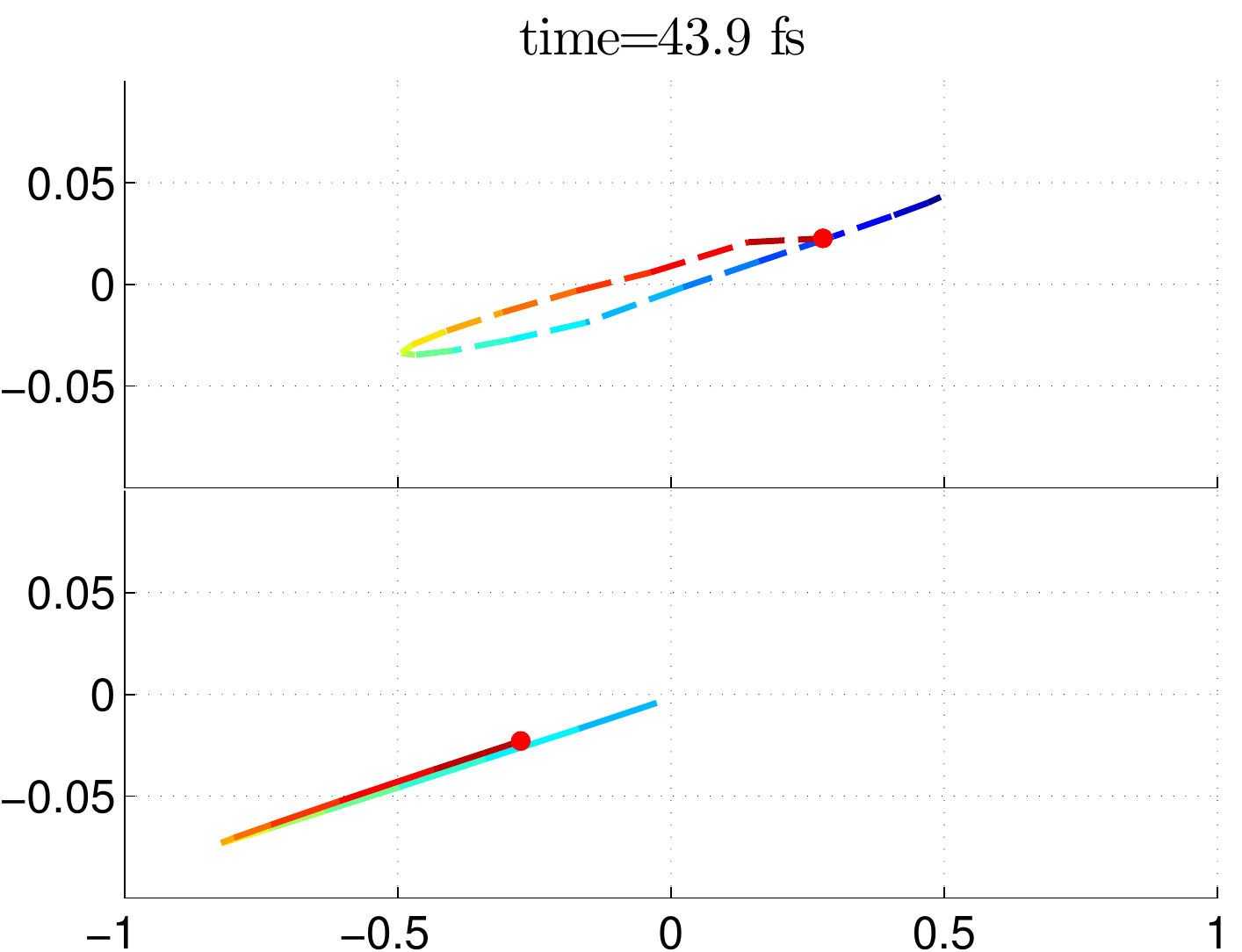}}&
      \resizebox{65mm}{!}{\includegraphics{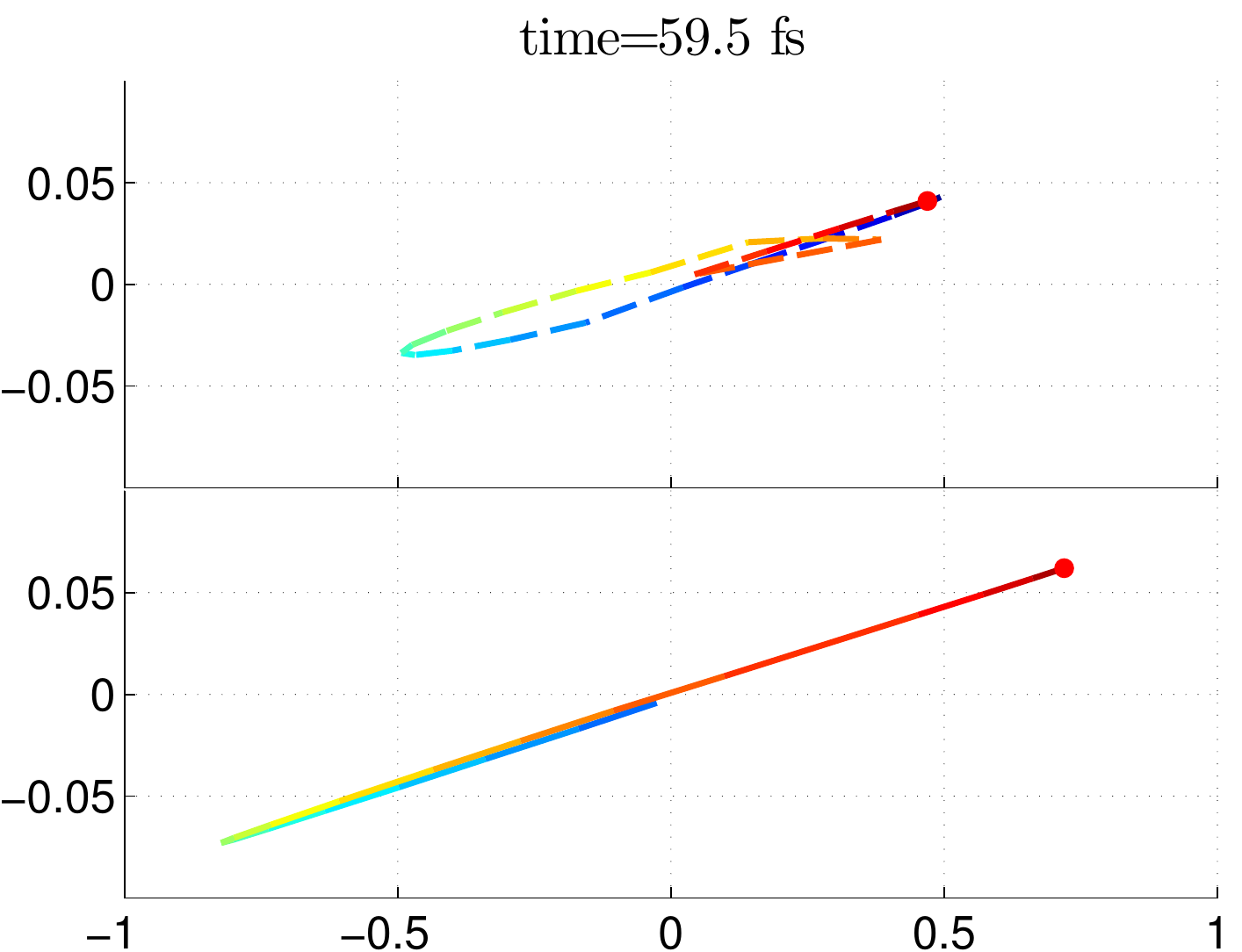}} \\
       \resizebox{65mm}{!}{\includegraphics{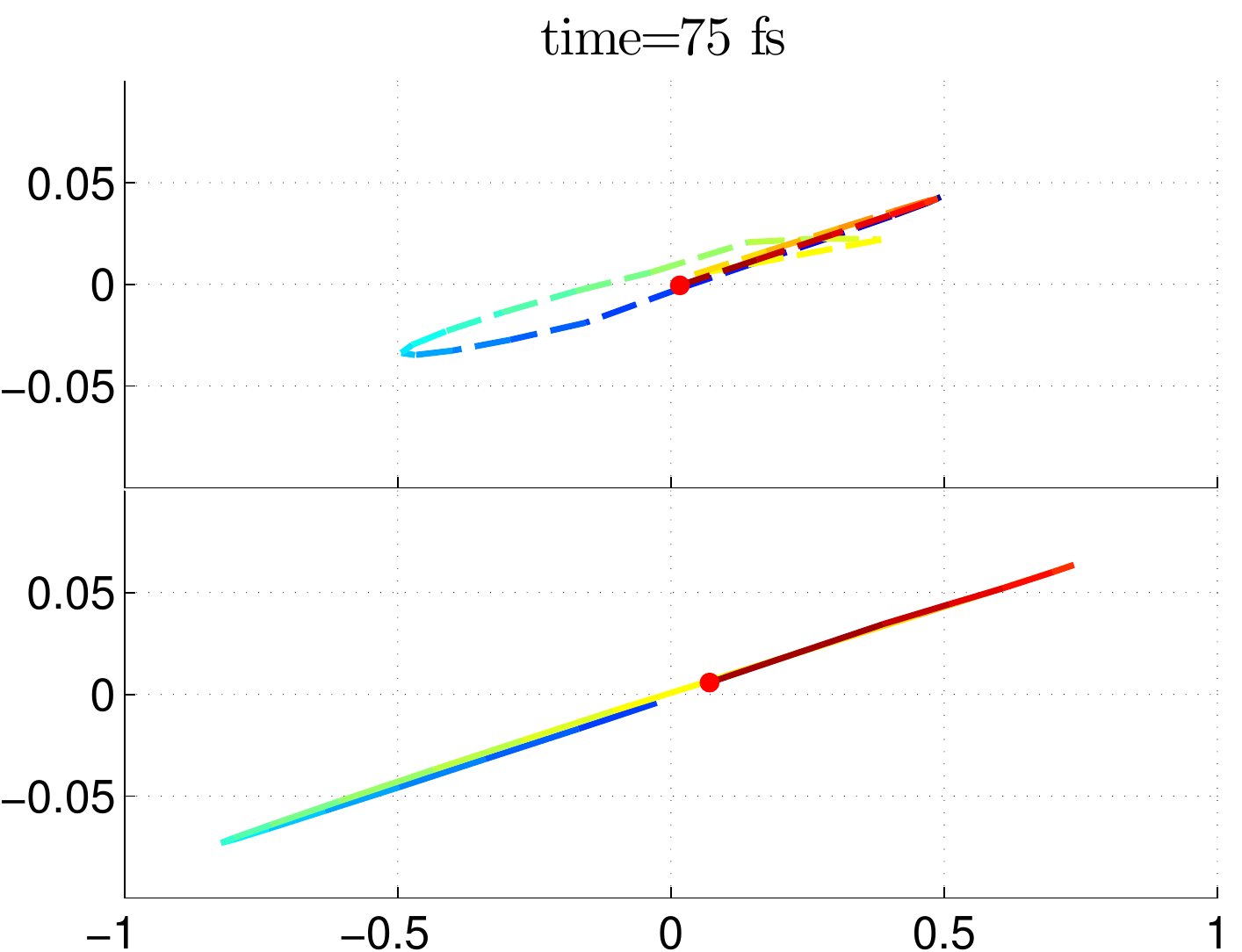}} &
      \resizebox{65mm}{!}{\includegraphics{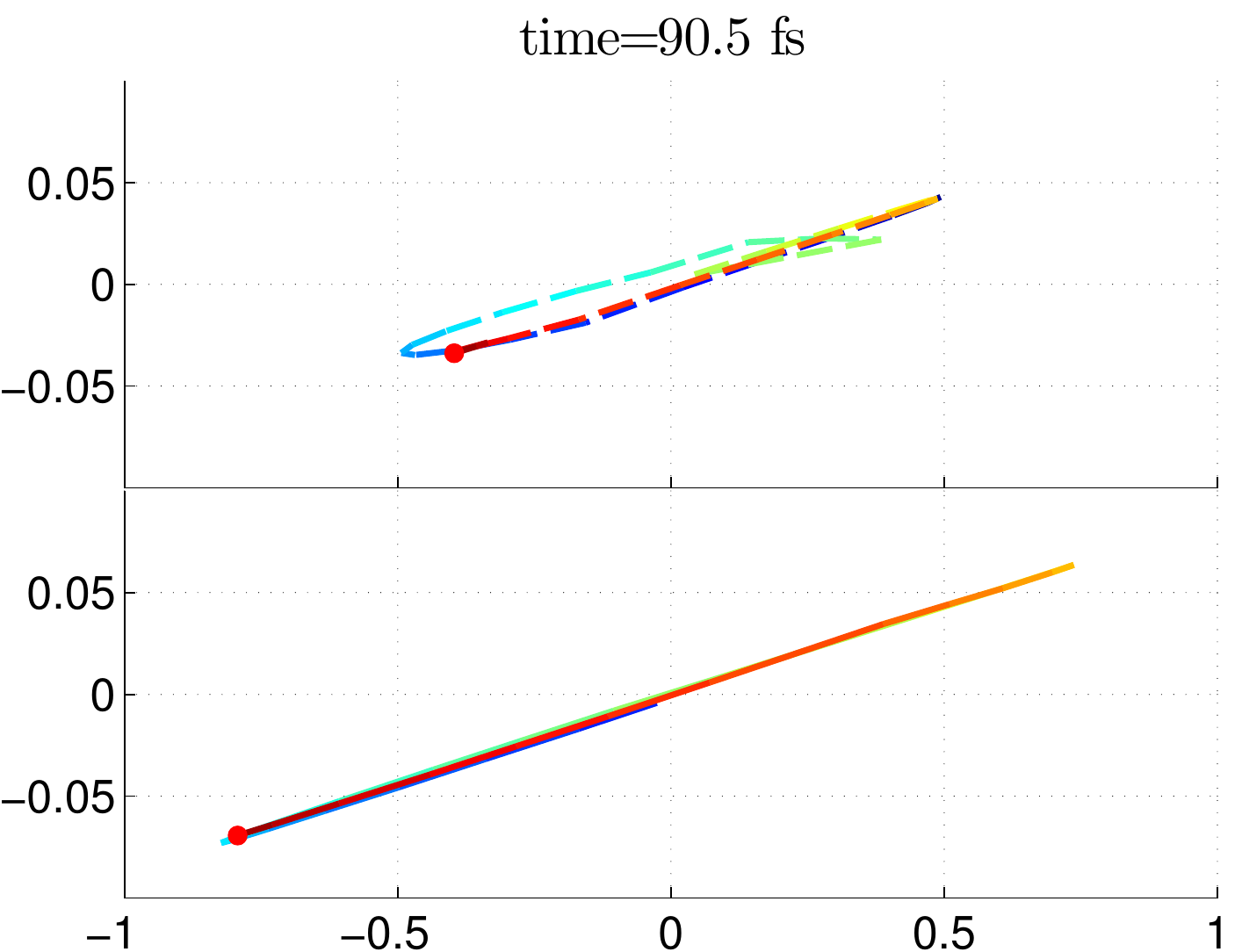}} \\
       \resizebox{65mm}{!}{\includegraphics{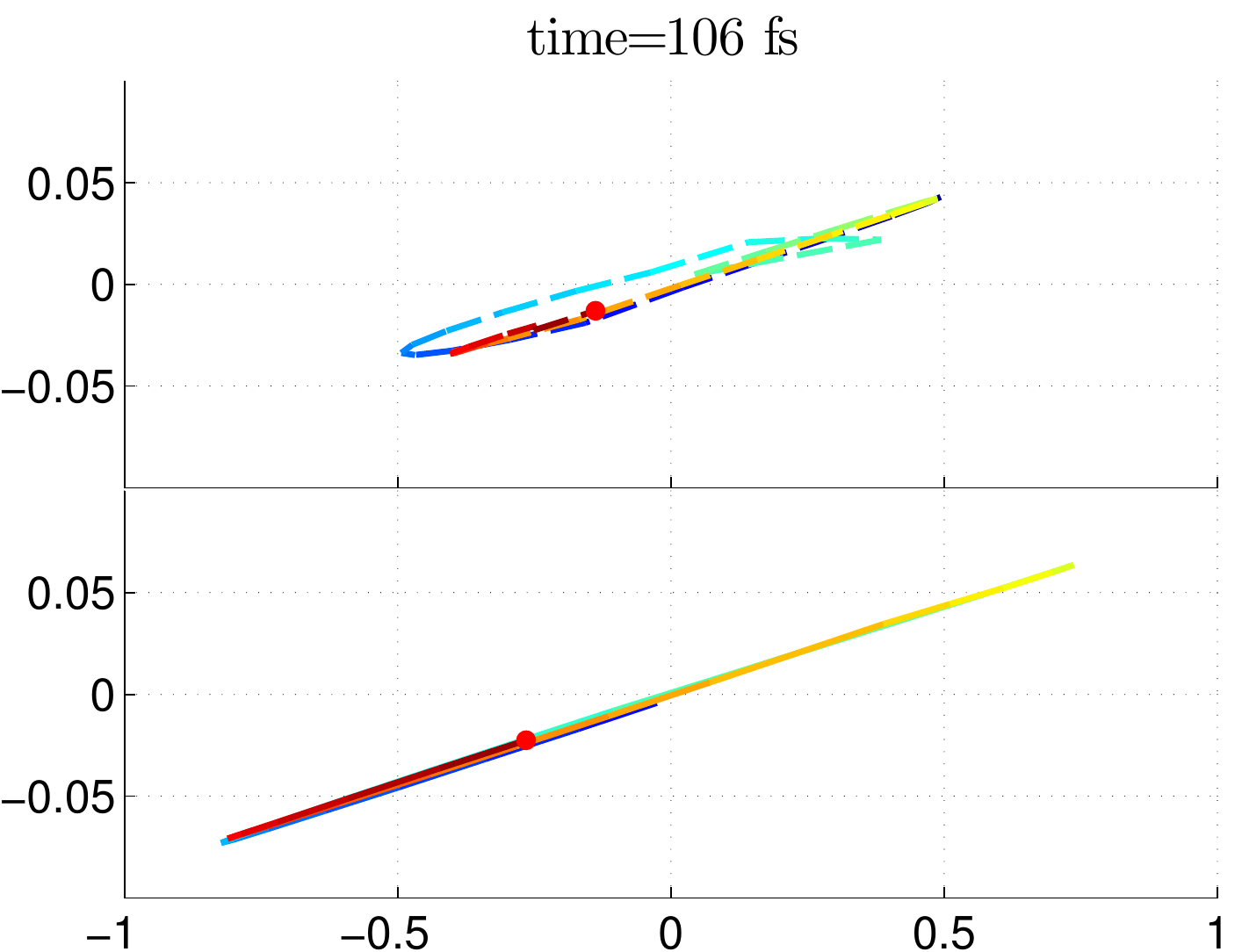}} &
      \resizebox{65mm}{!}{\includegraphics{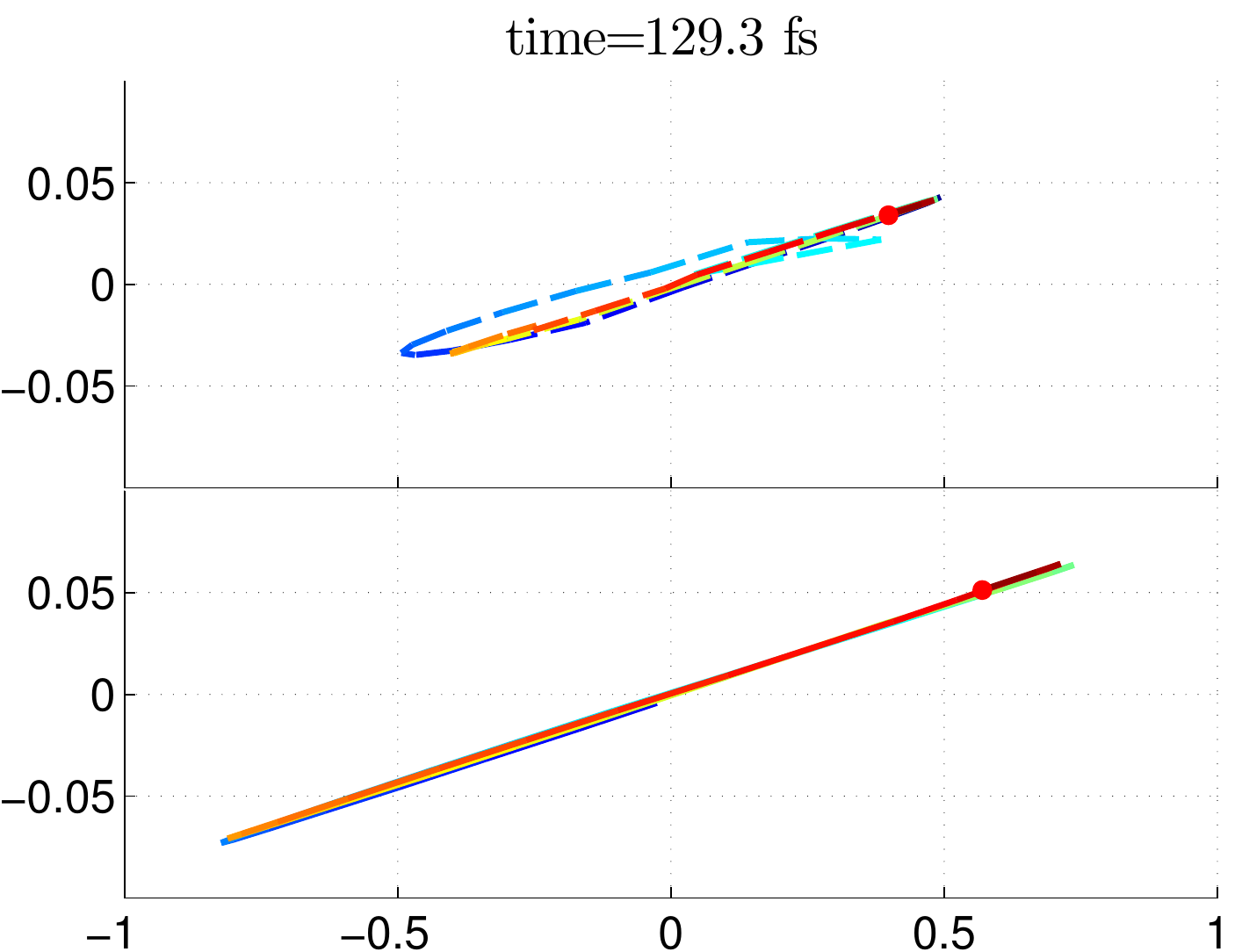}} \\
     \end{tabular}
    \caption{Average position for the upper and the lower surfaces at different times. The coordinates axis of each panel are $q_{1}$ and $q_{2}$. The red marker represents the average position, while the coloured curve shows positions visited at earlier times. The colour code is such, that positions visited more recently are in red compared to those visited at earlier times appearing in blue.}
    \label{fig:AvPos}
\end{figure*}

In Fig.~\ref{fig:AvPos} we show the expected position as a function of time. Ignoring nonadiabatic effects, 
one might expect that the average position of the wave packet follows a straight line through the conical intersection, since the initial momentum expectation equals zero, and the potential energy surfaces are radially symmetric. However, this expectation is not met, which can be explained by the surface hopping approximation: 
during the time interval $[17 {\rm fs}, 51 {\rm fs}]$, the wave packet is almost entirely located on the lower surface. In this case, the few trajectories on the upper surface, initially sampled from the tail of the Wigner distribution, gain relative weight with respect to the trajectories that have initiated nonadiabatic transitions. 
Because points sampled from the tail of the distribution are more likely to be arranged in a non symmetric way with respect to the origin in momentum space,
the average momentum does not point into the direction of the conical intersection.

In Fig.~\ref{fig:Error} we show the absolute deviation with respect to the reference solution for the expectation values of position, momentum and population referal to the upper surface. The differences are larger when the wave function is mostly located on the lower surface namely, for the time intervals $[17 {\rm fs},51 {\rm fs}]$ and $[77 {\rm fs},108 {\rm fs}]$. In particular, during the second time interval, the deviation on the three observables is amplified due to interference effects. The wave packet relative to the upper and lower surface arrive simultaneously at the conical intersection (see Fig.~\ref{fig:AvPos} at $\mbox{time}=75 \mbox{ fs}$). It is also important to notice that the curves describing the difference of the two Landau-Zener transition probabilities overlap quite well on the scale of the deviation with respect to the reference solution; hence, our experiments confirm the closeness of the two LZ probabilites for small values of the gap.

\begin{figure}[ht!]
  \begin{center}
      \resizebox{85mm}{!}{\includegraphics{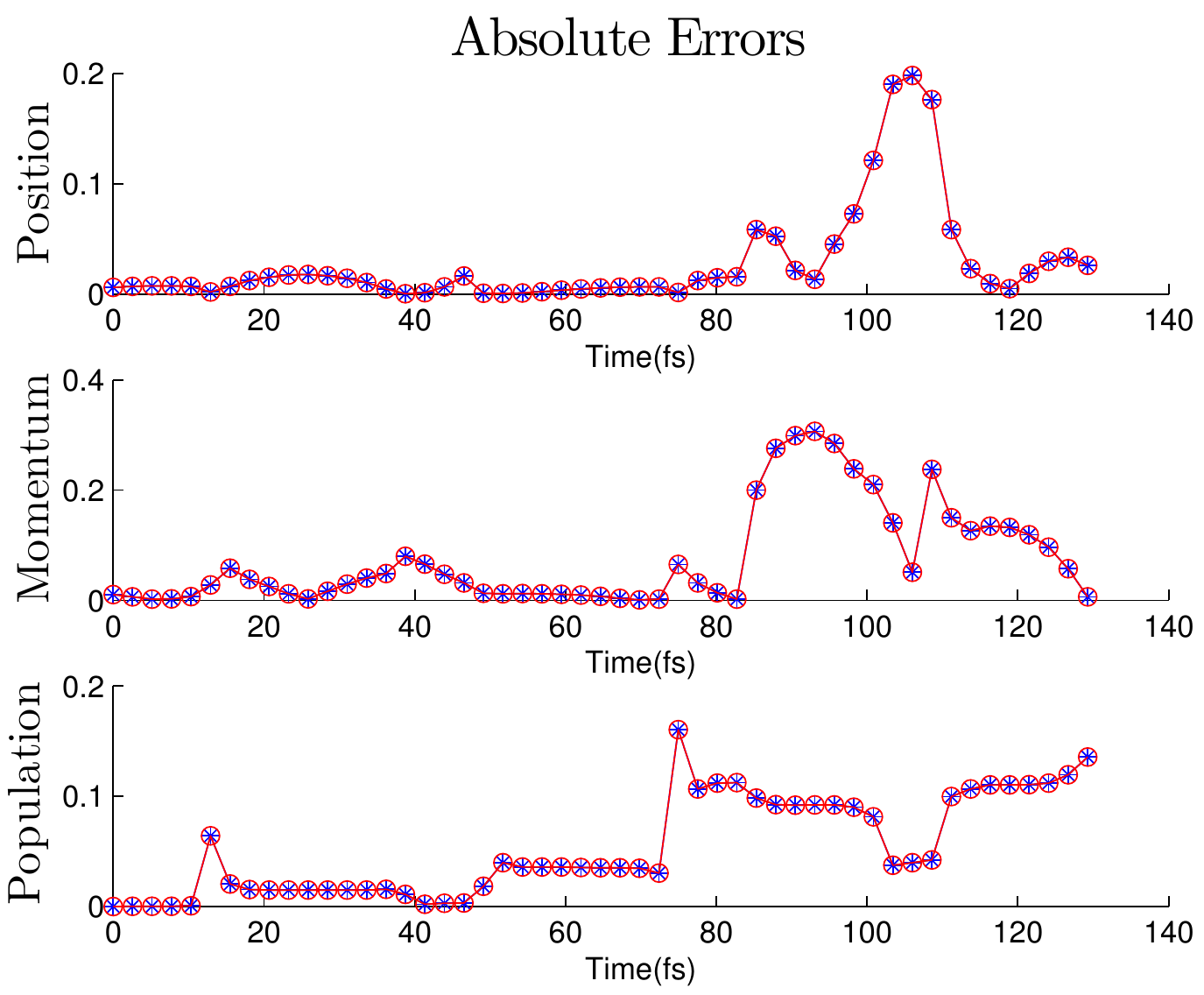}} 
    \caption{Absolute deviation of the first component of the position expectation (upper panel), momentum expectation (middle panel) and population (lower panel) of the upper surface with respect to the reference solution. 
    The blue markers are relative to the simulation obtained using the diabatic LZ formula, those in red to the adiabatic one.}
    \label{fig:Error}
  \end{center}
\end{figure}

\subsection{LZ probabilities}
\begin{figure}[ht!]
  \begin{center}
    \resizebox{85mm}{!}{\includegraphics{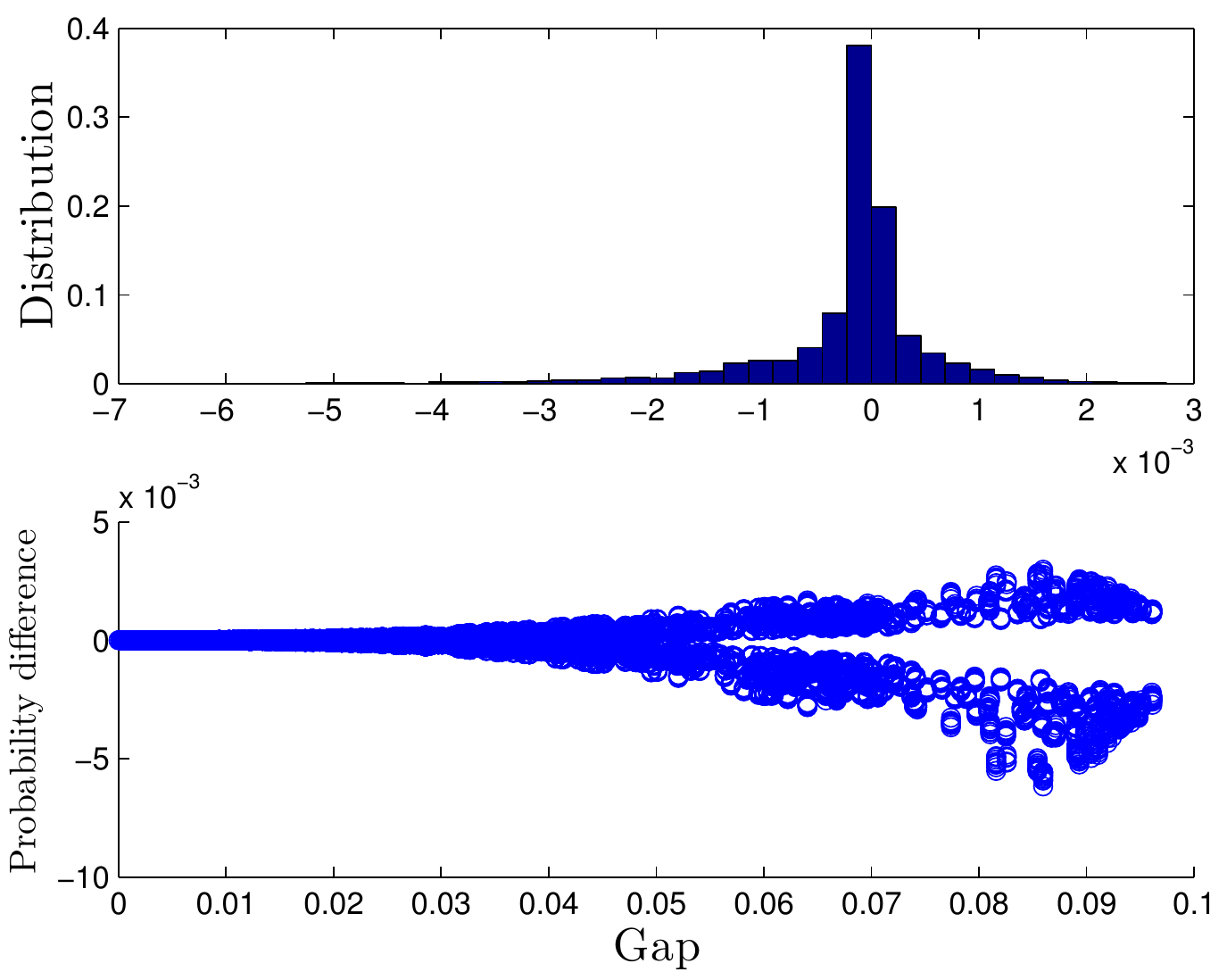}}
    \caption{Difference between the transition probabilities. Upper panel: distribution of $P^{\rm LZ}_{a}-P^{\rm LZ}_{d}$, the average value and standard deviation are $\mu=3.82\times10^{-4}$ and $\sigma=1.2\times 10^{-3}$, respectively. Lower panel: $P^{\rm LZ}_{a}-P^{\rm LZ}_{d}$ vs the gap value $Z(q_c)$.}
    \label{fig:LZProb}
  \end{center}
\end{figure}

In Fig.~\ref{fig:LZProb}, we analyze the difference between the LZ probabilities (\ref{eq:dia}) and (\ref{eq:ad}) computed simultaneously for each trajectory at each local minimum of the gap. In particular, we notice that the magnitude of the difference between the two transition probabilities is mostly $10^{-3}$ or smaller and increases up to $5 \times 10^{-3}$ as the value of the gap function increases.
The two branches shown in the lower panel of Fig.~\ref{fig:LZProb} are explained by the specific form of LZ probabilities for the linear Jahn--Teller case: We have  
$$
P^{\rm LZ}_{d}=\exp\left(-\frac{\pi}{\eps}\frac{|q_c|^{2}}{|p_c^\pm|}\right),
$$
and obtain by the calculation of section \S\ref{sec:DifferentLZ}
\begin{equation}\label{eq:PaJT}
P^{\rm LZ}_{a}=\exp\left(-\frac{\pi}{\eps}\frac{|q_c|^{2}}{\sqrt{|p_c^\pm|^{2}-2\gamma|q_c|^{2}\mp |q_c|}}\right),
\end{equation}
where the plus and minus sign refer respectively to transitions from the upper level to the lower and vice versa. 

In the linear Jahn--Teller situation, we have $P^{\rm LZ}_{d} < P^{\rm LZ}_{a}$ for transitions from the upper to the lower surface and $P^{\rm LZ}_{d} > P^{\rm LZ}_{a}$ for transitions from the lower to the upper surface, which explains the two branches in the lower panel.

\begin{figure}[!ht]
  \begin{center}
      \resizebox{85mm}{!}{\includegraphics{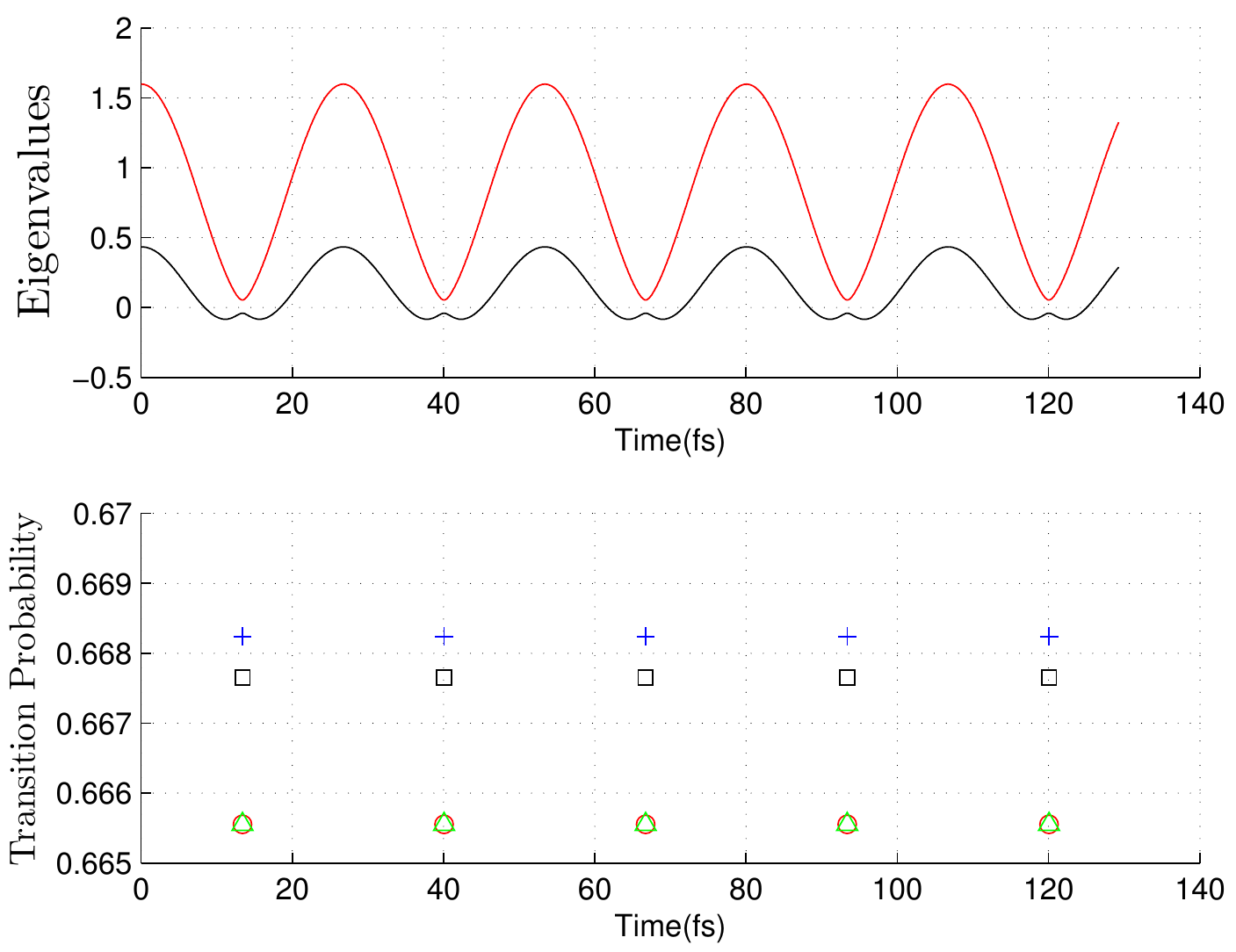}}     
    \caption{Eigenvalues and transition probabilities of a trajectory located on the upper surface when using the deterministic method. Upper panel: eigenvalues relative to the upper level (red) and lower level (black), respectively. Lower panel: transition probabilities in correspondence to the local minima of $t\mapsto Z(q^+(t))$; the red circles refer to the adiabatic LZ probability, while the blue crosses refer to the diabatic one. Black and green markers represent formula (\ref{eq:PaJT}) and $P^{\rm LZ}_0$ respectively. The slight difference between the red and green markers is due to the numerical error when computing the second derivative of $t\mapsto Z(q^+(t))$.}
      \label{fig:Eigenvalues1}
  \end{center}
\end{figure}

\begin{figure}[!ht]
  \begin{center}
      \resizebox{85mm}{!}{\includegraphics{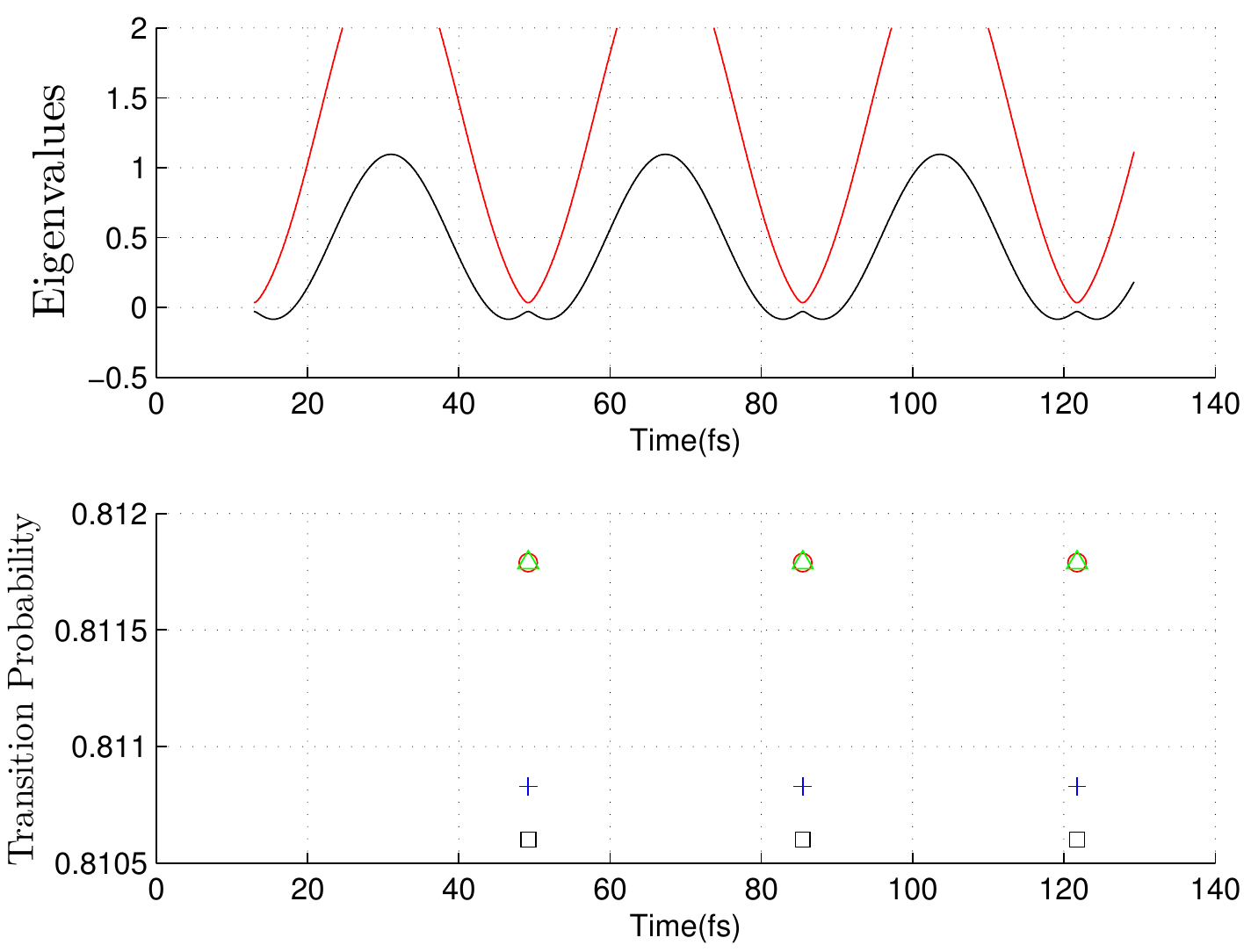}}     
    \caption{Eigenvalues and transition probabilities along a trajectory located on the lower surface when using the deterministic method. The colors and markers correspond to the ones in Fig.~\ref{fig:Eigenvalues1}.}
    \label{fig:Eigenvalues2}
  \end{center}
\end{figure}

Next we monitor individual trajectories located on the lower and the upper surface respectively. The upper panel of Fig.~\ref{fig:Eigenvalues1} represents the values of the two eigenvalues $U^{+}$ and $U^{-}$ along a typical upper surface trajectory. For each local minimum of $t\mapsto Z(q^+(t))$ we compute the transition probability in four different ways: We use the diabatic and the adiabatic formulas (\ref{eq:dia}) and (\ref{eq:ad}) respectively, the Jahn--Teller specific analytic version of the adiabatic formula (\ref{eq:PaJT}) and the intermediate probability 
$$
P^{\rm LZ}_{0} = \exp\left(-\frac{\pi}{\eps}\frac{|q_c|^{2}}{\sqrt{|p_c^\pm|^{2}-2\gamma|q_c|^{2}}}\right),
$$
which lacks the surface dependent term $\mp|q_c|$ of (\ref{eq:PaJT}). In Fig.~\ref{fig:Eigenvalues2}, we show the corresponding information for a typical lower level trajectory.

Figs.~\ref{fig:Eigenvalues1} and \ref{fig:Eigenvalues2} depict that the differences between transition probabilities calculated by means of the different LZ formulas are small. It is worth emphasizing that these deviations are within the accuracy of the LZ approximation, since the conventional LZ formula is obtained assuming a constant and high momentum, that is, $|p_c|^2 \gg Z(q_c)=2|q_c|$. The formulas derived in the present section for the linear Jahn-Teller case clearly show that the diabatic and the adiabatic formulas coincide in the high-energy regime, while the corrections are mainly due to an acceleration and are of the order of the energy gap. 

Moreover, Figs.~\ref{fig:Eigenvalues1} and \ref{fig:Eigenvalues2} also show that local nonadiabatic regions along classical trajectories have the form of an avoided crossing, although the global nonadiabatic region is formed by conically intersecting adiabatic energy surfaces. 
Since classical trajectories passing exactly through a conical intersection point are very rare, almost all local nonadiabatic regions along trajectories have the form of one-dimensional avoided crossings. 
The adiabatic LZ formula (\ref{eq:ad}) has been applied to the one-dimensional case of several avoided crossings in atomic Na+H collisions\cite{BelyaevLebedev:2011}, and a good approximation of the quantum results has been found. 
The same is true for the diabatic formula~(\ref{eq:dia}), which has also been successfully applied to one-dimensional avoided crossings\cite{FermanianLasser:2012}.


The above findings are also valid for different values of the semiclassical parameter $\eps$. In particular analogous distributions for the difference of $P^{\rm LZ}_{a}-P^{\rm LZ}_{d}$ are obtained when using $\eps=0.05$ ( mean and standard deviation being $\mu=9.78\times10^{-4}$ and $\sigma=1.3\times10^{-3}$ respectively) and $\eps=0.001$ ($\mu=4.57\times10^{-4}$ and $\sigma=1.3\times10^{-3}$).

\subsection{Probabilistic VS Deterministic}
In this section we compare the previous results with those obtained by the probabilistic approach. The simulations presented in this section are obtained taking the average over 10 runs with the same initial trajectories used for the deterministic approach.

\begin{figure}[!ht]
  \begin{center}
      \resizebox{85mm}{!}{\includegraphics{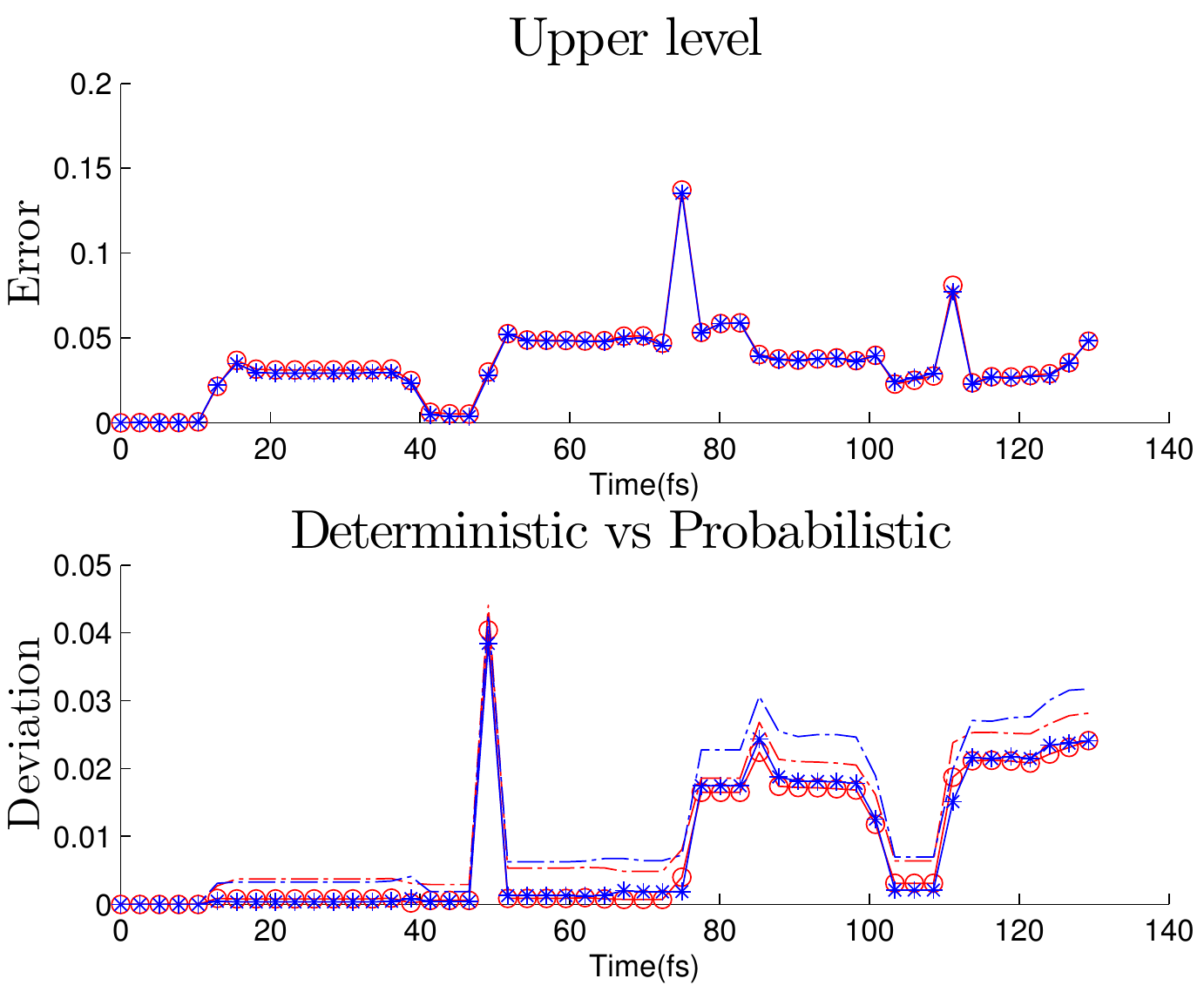}}     
    \caption{Upper panel: absolute error of the upper surface population with respect to the reference solution. The blue markers refer to the simulation obtained using $P^{\rm LZ}_{d}$, those in red to $P^{\rm LZ}_{a}$. Lower panel: difference, in the absolute value, of the level population between the deterministic  and the probabilistic approach. The dotted line indicates the confidence interval.}
    \label{fig:Popdet}
  \end{center}
\end{figure}

The results obtained are analogous to those displayed in the deterministic case. In particular, in the lower panel of 
Fig.~\ref{fig:Popdet} we compare the population for the upper surface obtained by the deterministic and the probabilistic approach. 

\begin{figure*}[t]
    \begin{tabular}{llll}                                                                                      
      \resizebox{65mm}{!}{\includegraphics{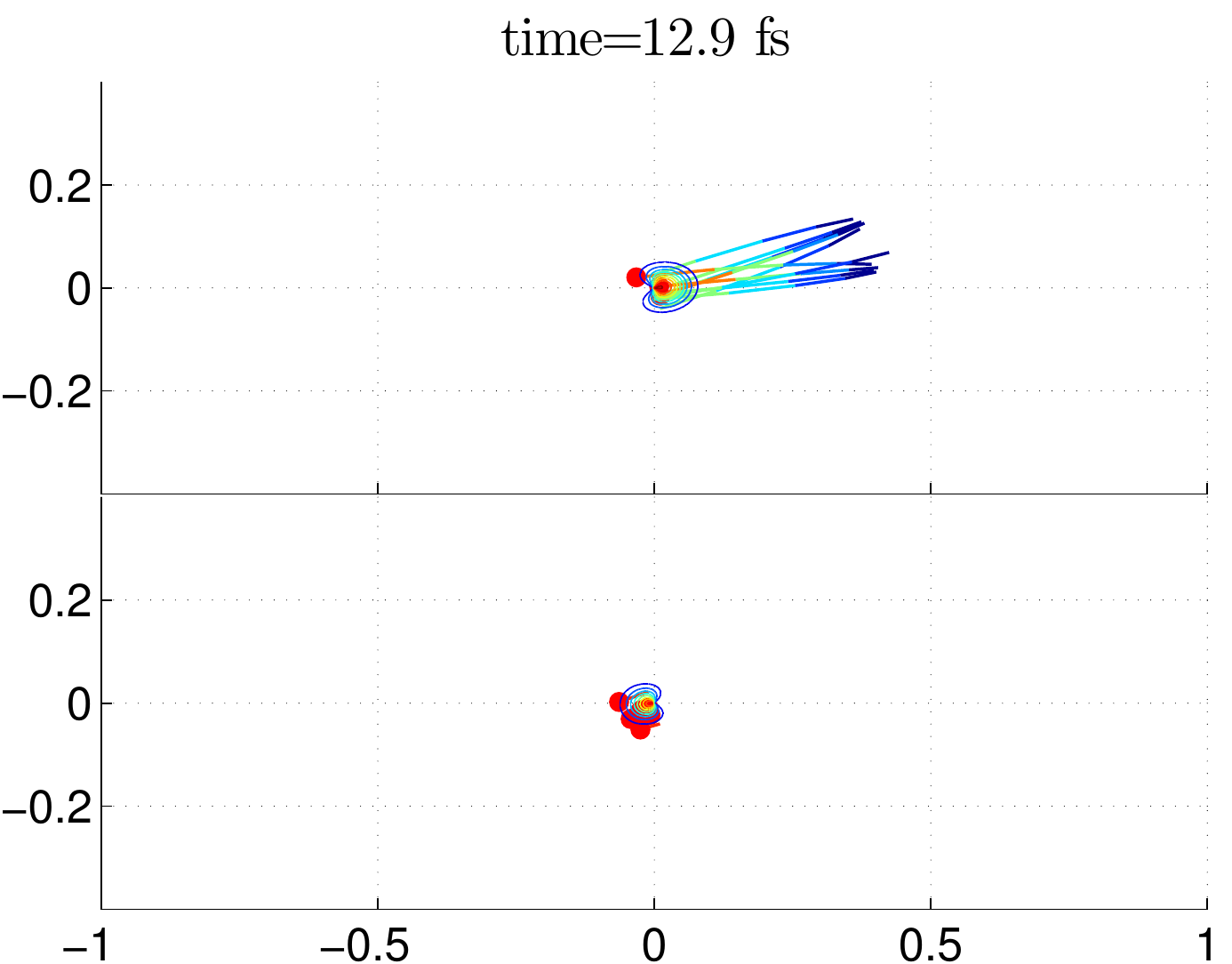}} &
      \resizebox{65mm}{!}{\includegraphics{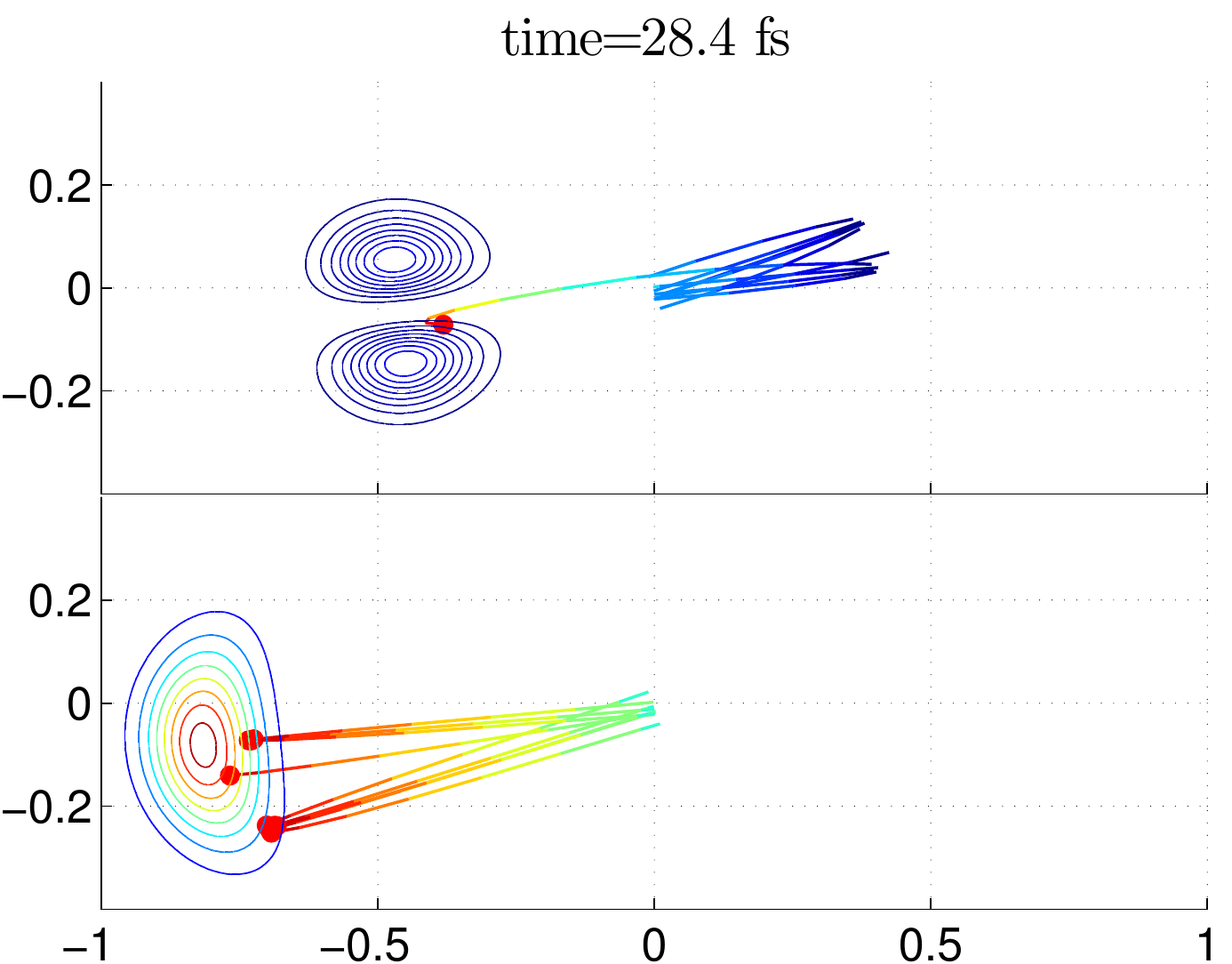}} \\
       \resizebox{65mm}{!}{\includegraphics{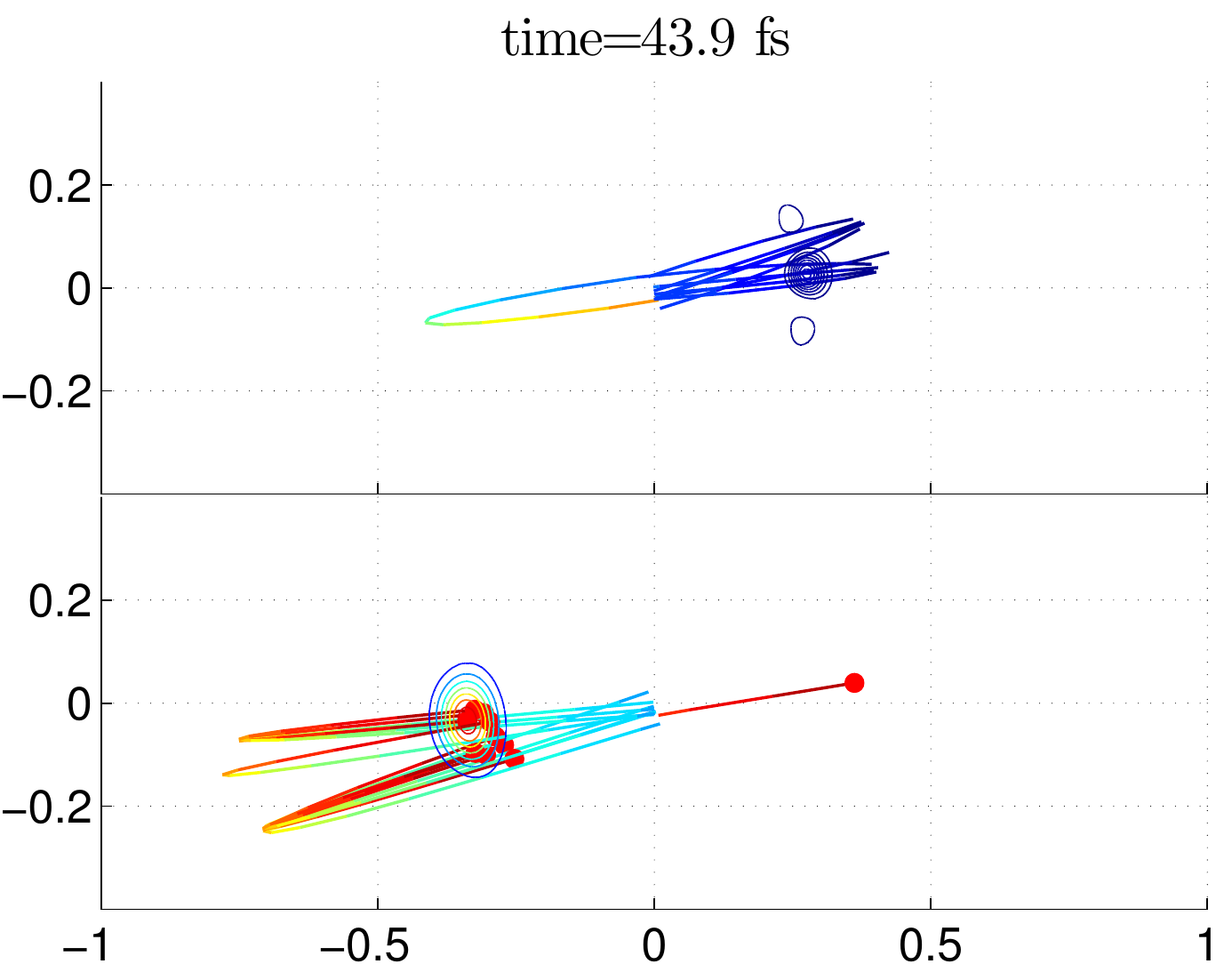}}&
      \resizebox{65mm}{!}{\includegraphics{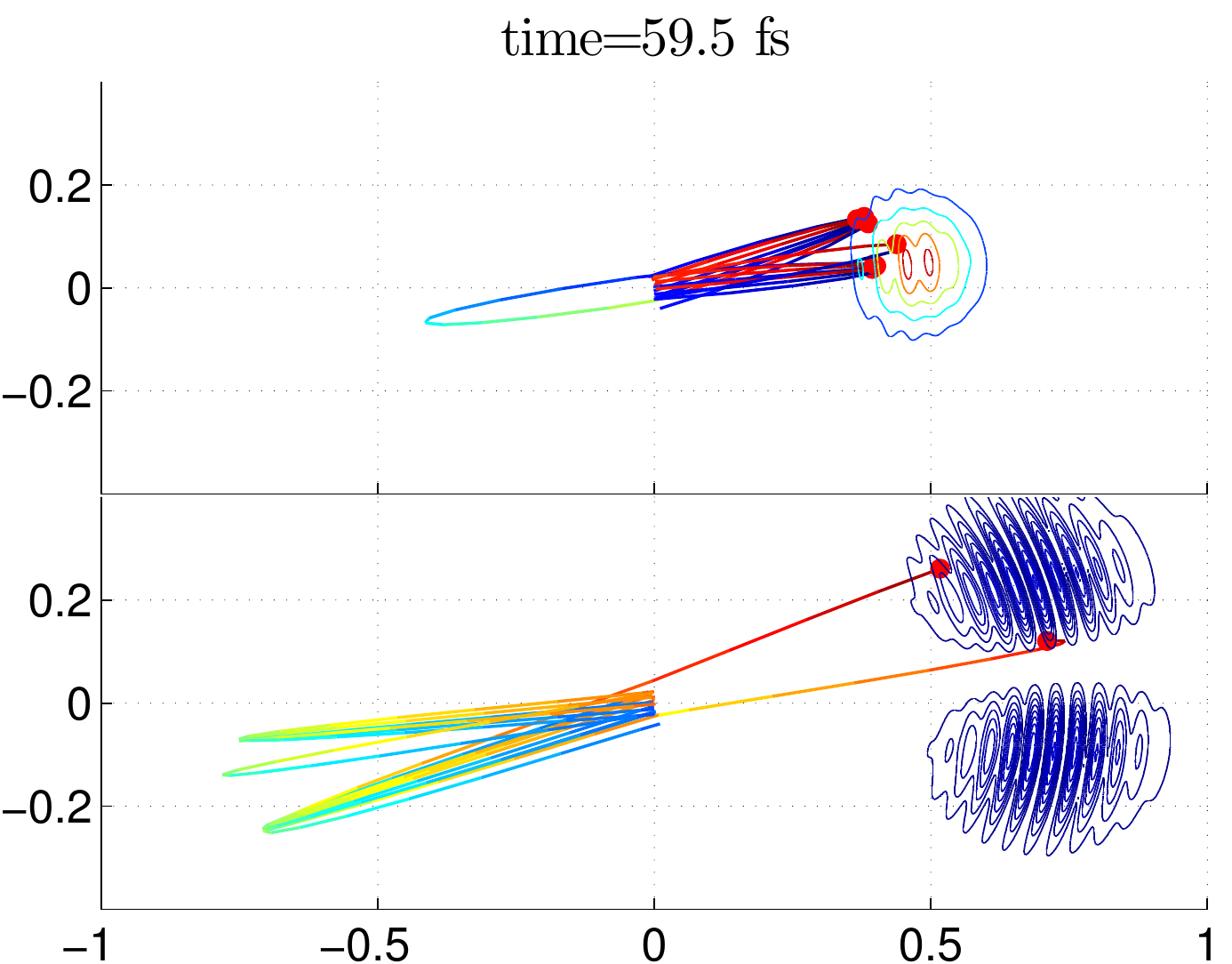}} \\
       \resizebox{65mm}{!}{\includegraphics{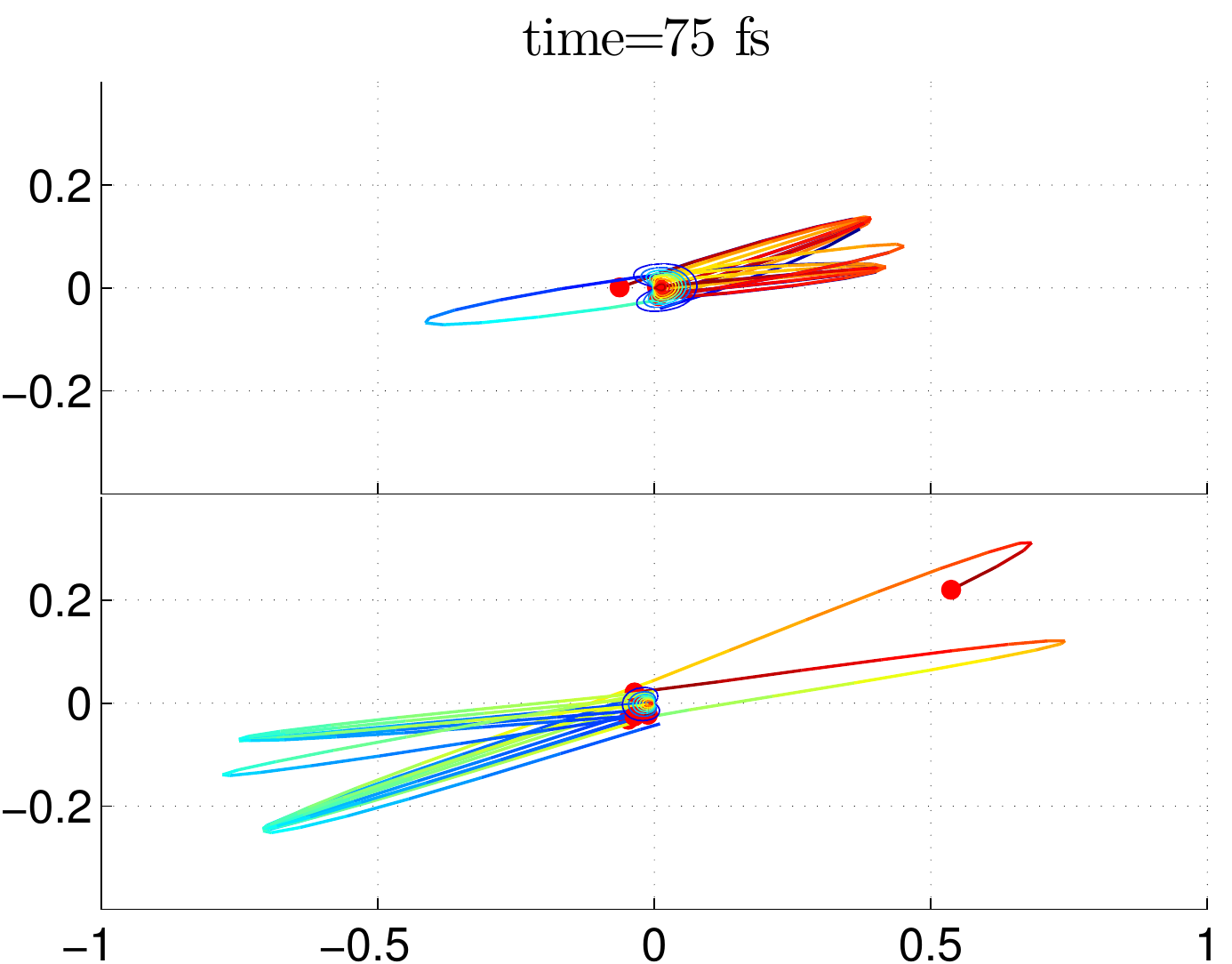}} &
      \resizebox{65mm}{!}{\includegraphics{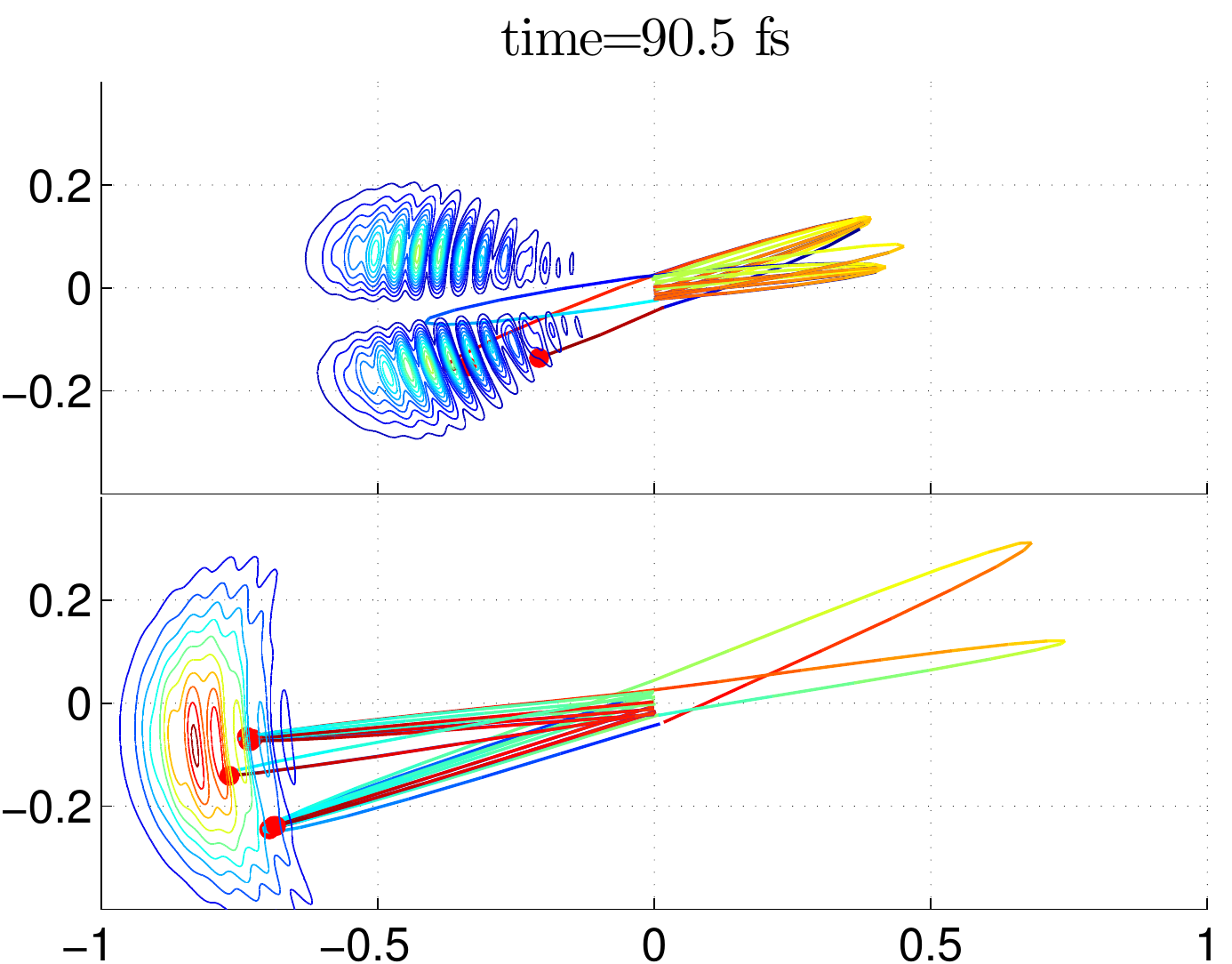}} \\
       \resizebox{65mm}{!}{\includegraphics{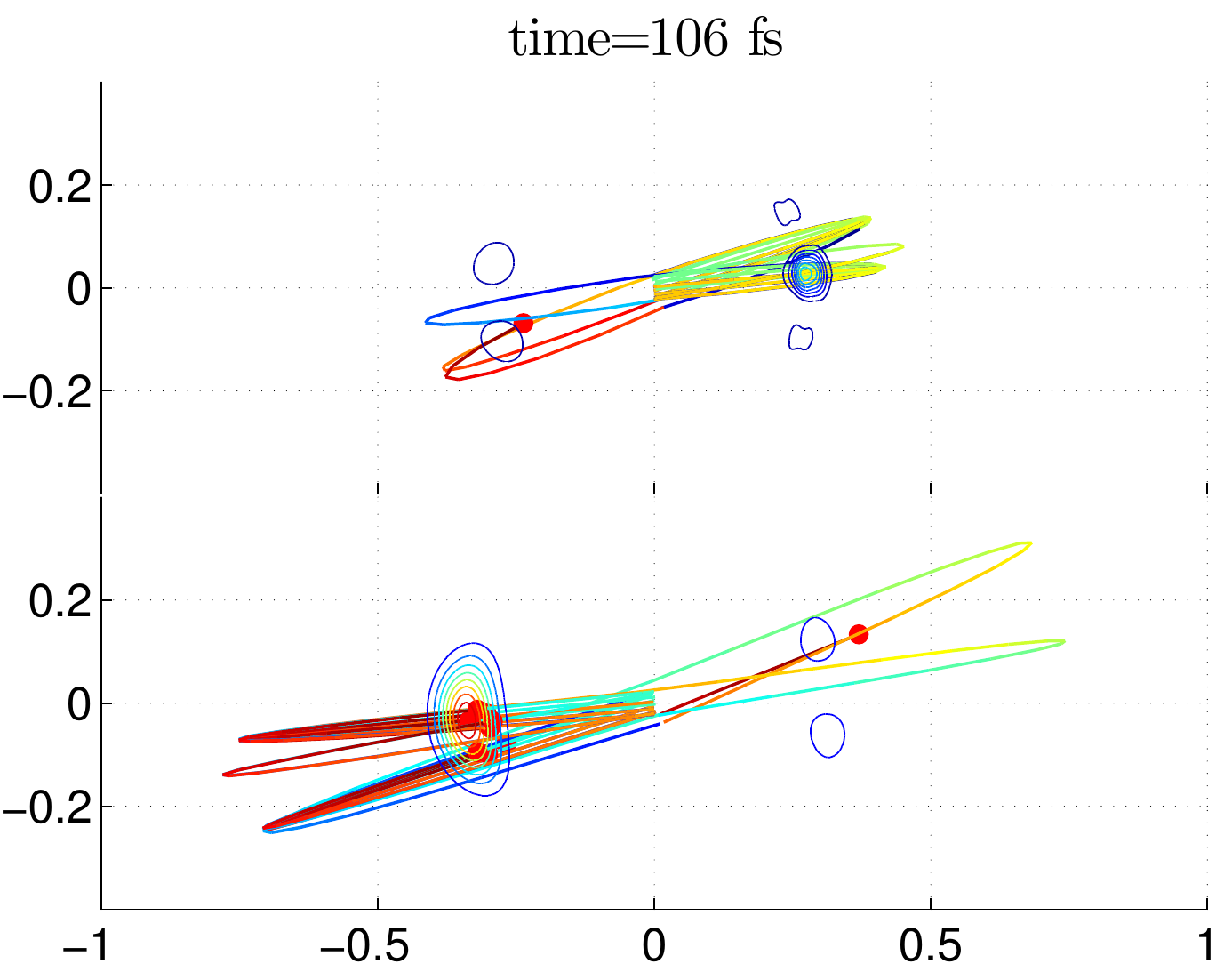}} &
      \resizebox{65mm}{!}{\includegraphics{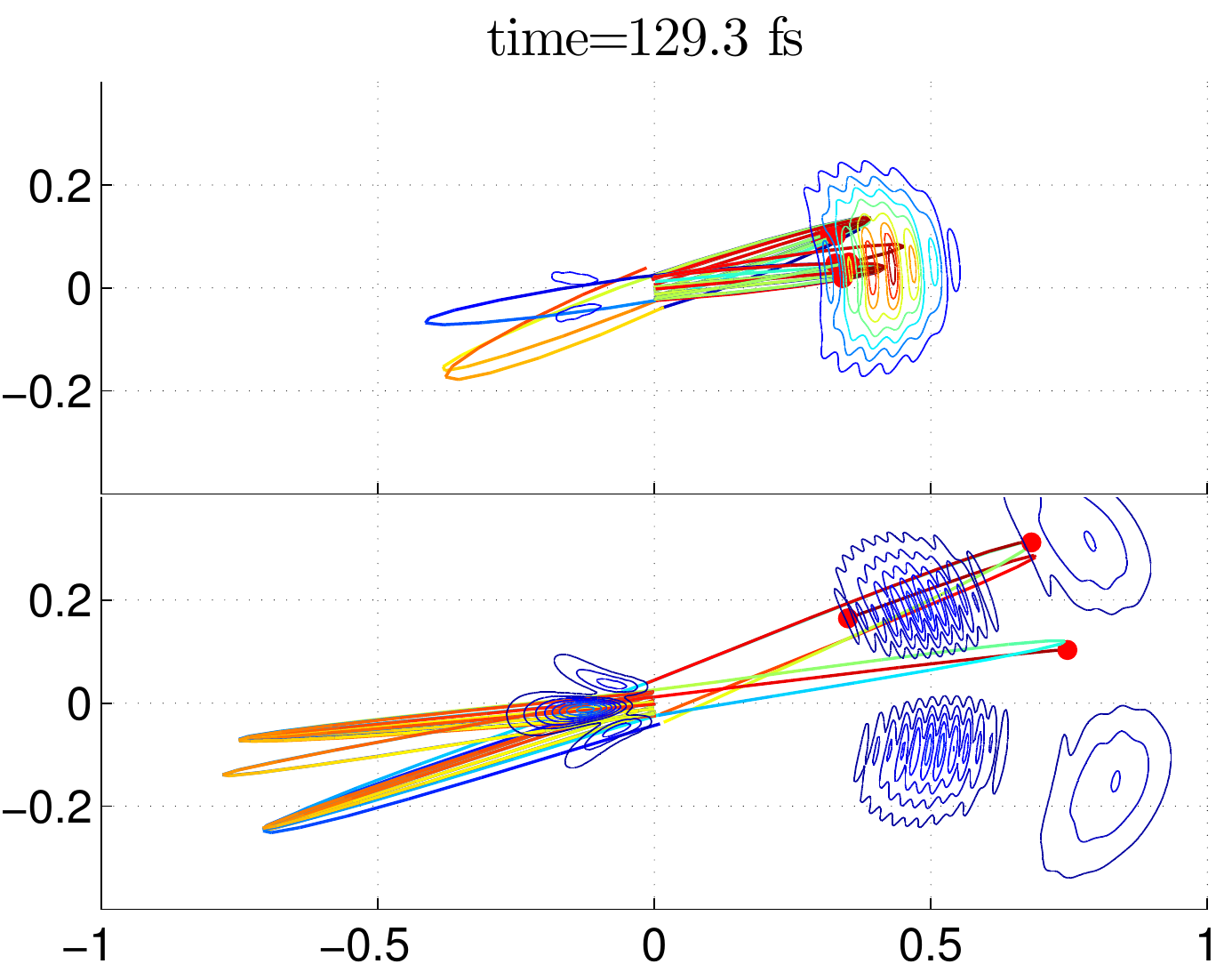}} \\
     \end{tabular}
    \caption{Sample of trajectories in the probabilistic setting together with the modulus of the projected wave function obtained by solving the Schr\"{o}dinger equation. Each panel of the figure refers to the upper and lower surface at a given time. The trajectories are represented with a colored curve ending with a red marker representing the position at the time indicated on each panel. The colour code is so that positions visited more recently are in red compared to those appearing in blue referring to positions visited earlier in the past.}
    \label{fig:Traj}
\end{figure*}

The final Fig.~\ref{fig:Traj} shows the time evolution of a sample of typical surface hopping trajectories. As expected, their positions fit with the position density of the reference solution.  

Comparing the deterministic and probabilistic approaches, we point out that probability currents computed by a quantum method split when 
passing through nonadiabatic regions, see e.g. Fig.~5 of Ref.\cite{BelyaevGrosser:1996}. The deterministic approach with its branching classical trajectories simulates this situation. Moreover, the deterministc method has been mathematically analysed\cite{FermanianLasser:2008}. On the other hand, in some cases the deterministic approach may produce too many trajectories, so that memory requirements and computing times become unfeasible and the probabilistic method with its constant number of trajectories is preferable.

\section{Conclusion}
We have investigated a class of surface hopping algorithms, which perform nonadiabatic transitions for each classical trajectory individually. Nonadiabatic transitions are allowed, when the surface gap attains a local minimum along an individual trajectory. We have compared two recent Landau--Zener formulas for the probability of nonadiabatic transitions, one of them requiring a diabatic representation of the potential matrix, the other one only depending on the adiabatic potential energy surfaces. Our numerical experiments confirm the expected affinity of both LZ 
probabilities as well as the good approximation of reference values, that have been obtained by a grid based quantum solver. We have visualized position expectations and superimposed surface hopping trajectories with reference position densities for an enhanced understanding of the effective dynamics. 

\begin{acknowledgements}
AKB gratefully acknowledges supports from the Russian Foundation for Basic Research (Grant No. 13-03-00163-a). 
All three authors have been supported by the German Research Foundation (DFG), Collaborative Research
Center SFB-TRR 109. We thank Diane Clayton-Winter for her reading of the manuscript.
\end{acknowledgements}

\end{document}